\documentclass[aps,amsmath,amssymb,twocolumn]{revtex4-2}

\usepackage{amsmath} 
\usepackage{amssymb} 
\usepackage{graphicx}
\usepackage{dcolumn}
\usepackage{bm}
\usepackage{color}
\usepackage[normalem]{ulem}
\usepackage{algpseudocode}
\usepackage[linesnumbered,ruled,vlined]{algorithm2e} 
\DeclareMathOperator*{\argmax}{arg\,max}
\usepackage{dcolumn}  % Contains a mode for making columns in tables
\usepackage{hyperref}% add hypertext capabilities
\newcolumntype{C}[1]{>{\centering\arraybackslash}p{#1}}
\newcolumntype{L}[1]{>{\raggedright\arraybackslash}p{#1}}

\begin{document}

\title{Deep Reinforcement Learning for Radiative Heat Transfer Optimization Problems}

\author{E. Ortiz-Mansilla$^{\dagger}$}
\author{J.~J. Garc\'{\i}a-Esteban$^{\dagger}$}
\author{J. Bravo-Abad}
\author{J.~C. Cuevas}
\email{juancarlos.cuevas@uam.es}

\affiliation{Departamento de F\'{\i}sica Te\'orica de la Materia Condensada, Universidad Aut\'onoma de Madrid,
28049 Madrid, Spain}

\affiliation{Condensed Matter Physics Center (IFIMAC), Universidad Aut\'onoma de Madrid, 28049 Madrid, Spain}

\date{\today}

\begin{abstract}
Reinforcement learning is a subfield of machine learning that is having a huge impact in the different 
conventional disciplines, including physical sciences. Here, we show how reinforcement learning methods can be 
applied to solve optimization problems in the context of radiative heat transfer. We illustrate their use with the 
optimization of the near-field radiative heat transfer between multilayer hyperbolic metamaterials. Specifically, we 
show how this problem can be formulated in the language of reinforcement learning and tackled with a variety of 
algorithms. We show that these algorithms allow us to find solutions that outperform those obtained using physical 
intuition. Overall, our work shows the power and potential of reinforcement learning methods for the 
investigation of a wide variety of problems in the context of radiative heat transfer and related topics.
\end{abstract}

\maketitle

\section{Introduction}

Thermal radiation is an ubiquitous physical phenomenon whose understanding is of critical importance for many different 
areas of science and engineering \cite{Modest2013,Howell2020,Zhang2020}. The field of radiative heat transfer is enjoying 
a revival due to various recent advances \cite{Cuevas2018}. Maybe the most notable one is the demonstration that
the near-field radiative heat transfer (NFRHT) between two closely placed bodies can largely overcome the blackbody limit 
set by Stefan-Boltzmann's law. This was predicted in the early 1970s \cite{Polder1971Nov} and it has been verified in recent 
years in a large variety of systems with the help of novel experimental techniques \cite{Song2015,Cuevas2018,Biehs2021}. 
This effect originates from the fact that, when two objects are separated by a distance smaller than the thermal wavelength 
$\lambda_{\rm Th}$ ($\sim$10 $\mu$m at room temperature), the radiative heat flux can be greatly enhanced by the additional 
contribution of evanescent waves -- which is not considered in Stefan-Boltzmann's law. Near-field thermal radiation has opened 
new possibilities and holds the promise to have a notable impact in different technologies such as heat-assisted magnetic 
recording \cite{Challener2009}, scanning thermal microscopy \cite{deWilde2006,Kittel2008,Jones2013}, coherent thermal 
sources \cite{Carminati1999,Greffet2002}, near-field based thermal management \cite{Cuevas2018,Biehs2021} or thermophotovoltaics
\cite{Mittapally2023}.

NFRHT is by no means the only breakthrough in the field of thermal radiation in recent times. Thus, for instance, it has been 
shown that nanophotonic structures, where at least one of the structural features is at subwavelength scale, can have 
thermal radiation properties that differ drastically from those of conventional thermal emitters~\cite{Li2018}.
This has led to the development and improvement of energy applications such as daytime passive radiative 
cooling~\cite{Rephaeli2013,Raman2014}, thermal radiative textiles~\cite{Tong2015,Hsu2016}, radiative 
cooling of solar cells~\cite{Li2017}, or thermophotovoltaic cells~\cite{Lenert2014}. On a more fundamental level, another 
remarkable discovery has been the possibility of overcoming the far-field limits set by Planck's law in the context of 
the thermal emission and the radiative heat transfer between subwavelength objects \cite{Biehs2016,Fernandez2018,Thompson2018Sep}.

At this stage, the physical mechanisms of radiative heat transfer in the different regimes are relatively well understood
and the interest is now shifting towards the optimization and design of novel thermal devices. This process is being
mainly assisted by physical intuition and standard numerical optimization methods. Thus, for instance, in the context 
of NFRHT, many different analytical upper bounds have been put forward to establish the limits of near-field thermal 
radiation \cite{Biehs2021,Chao2022}. These bounds are extremely ingenious, but often lack the ability to guide in practice 
the fabrication of actual structures. On the other hand, conventional numerical optimization techniques, such as Bayesian or
topology optimization \cite{Molesky2018}, are also being routinely used in the field. 

At the same time, the impressive achievements of machine learning techniques in different engineering areas have motivated 
many researchers to pursue a data-driven approach to investigate a plethora of problems in conventional science disciplines, 
including physical sciences \cite{Mehta2019,Carleo2019,Marquardt2021}. Radiative heat transfer is not an exception and in recent 
years different groups have applied various machine learning techniques to address key problems in this field. Most of the work 
thus far has been carried out with the help of artificial neural networks (ANNs) and deep learning algorithms. Thus for instance, 
we have shown how ANNs can be used to tackle optimization and inverse design problems in the context of NFRHT, passive 
radiative cooling and thermal emission of subwavelength objects \cite{Garcia-Esteban2021Dec}. There has also been a tremendous 
activity in the context of deep learning aided design and optimization of thermal metamaterials, for a recent review see 
Ref.~\cite{Zhu2024}. However, reinforcement learning, another subfield of machine learning, has been barely used in modern 
radiative heat transfer problems, with notable exceptions \cite{Yu2023}. Reinforcement learning (RL) is much closer to the 
layman's view of artificial intelligence and it deals with problems concerning sequential decision making \cite{SuttonBarto}. 
In RL, an agent learns via the interaction with an environment from which it receives feedback to make good decisions towards 
a given goal, such as the optimization of a physical process or the inverse design of a device.

In this work we want to fill this gap and show how RL can be used to tackle optimization problems in the context of
radiative heat transfer. To be precise, we illustrate the core ideas with a problem related to the optimization 
of NFRHT between multilayer hyperbolic metamaterials. We show how this type of problems can be framed in the language 
of RL and how different RL algorithms can be implemented to address them. In particular, we critically
assess the advantages and disadvantages of the different methods to help new users of RL to select the most convenient
algorithm for a given application. The methods presented in this work can be straightforwardly applied to a large variety 
of problems in the thermal radiation science and related fields.

The rest of the manuscript is organized as follows. In Sec.~\ref{sec-RL}, we briefly introduce the topic of RL
for non-experts to make our contribution more self-contained. In Sec.~\ref{sec-system}, we present the system and
problem that we have chosen to illustrate the use of RL in the context of thermal radiation problems, namely the 
optimization of NFRHT between multilayer hyperbolic metamaterials. Then, Sec.~\ref{sec-results} is devoted to 
the description of the main results of this work obtained with different RL algorithms. We have organized those
results according to the RL algorithm employed and we also provide a detailed description of such algorithms. 
Finally, we present some additional discussions and summarize our main conclusions in Sec.~\ref{sec-conclusions}.

\section{Reinforcement Learning: A brief reminder} \label{sec-RL}

In this section we provide a brief introduction to RL following Ref.~\cite{DeepRLbook}. This will allow us to set 
the language and make the manuscript more self-contained. Readers familiar with RL can safely skip this section.

RL is a subfield of machine learning that aims at solving sequential decision-making problems. Many 
problems can be formulated in this way, including those concerning the optimization of systems, devices, and 
processes in the physical sciences. To solve a problem within RL, we begin by defining a goal. Then, an 
algorithm takes actions and gets information about the external world on how well the goal is being achieved. 
To that end, we normally need to take many actions in a sequential fashion, where each action modifies the world 
around us. We observe the changes in the world and, with the help of the feedback we receive, we decide on the 
next action to take. 

\begin{figure}[t]
\includegraphics[width=0.85\columnwidth,clip]{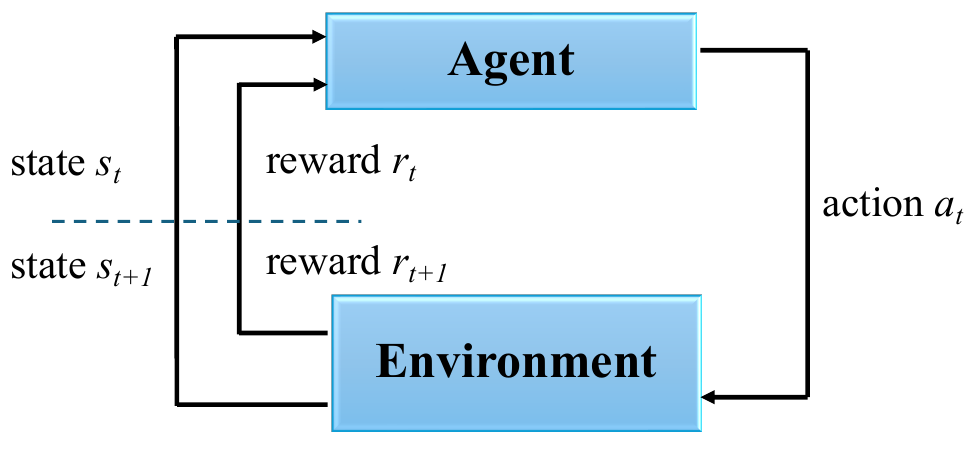}
\caption{The reinforcement learning control loop diagram.}
\label{fig-control-loop}
\end{figure}

The RL formulation of the process described above is the following. RL problems are formulated as a system that 
comprises an \emph{agent} and an \emph{environment}, the world surrounding the agent. The environment produces information 
which allows us to describe the \emph{state} of the system, while the agent interacts with the environment by observing 
the state and selecting an \emph{action}. The environment accepts the action and transitions into a new state, which is 
then observed by the agent to select a new action. In doing so, the environment also returns a \emph{reward} to the agent, 
which is used by the agent to select future actions. When the cycle of state $\rightarrow$ action $\rightarrow$ next state 
and reward is completed, we say that one \emph{time step} has passed. This cycle is repeated until the environment terminates, 
for example, when the problem is solved. This process is summarized in the control loop diagram of Fig.~\ref{fig-control-loop}.

Going deeper with the formulation of the process, a \emph{policy} in RL is the agent's action-producing function which maps
states to actions. As previously indicated, actions change the environment and affect what an agent observes and does next, 
which can be viewed as a sequential decision-making process which evolves in time. This RL process is driven by an 
\emph{objective}, which is defined as the sum of rewards received from the environment. The agent aims at maximizing the
objective by selecting good actions and learns to do this by interacting with the environment according to an optimizing 
policy in a trial-and-error process, which uses rewards to reinforce good actions and penalize bad ones. Therefore, the signals
exchanged between agent and environment are $(s_t, a_t, r_t)$, which stand for state, action, and reward, respectively, and 
where $t$ denotes the time step in which these signals occurred. The defined tuple $(s_t, a_t, r_t)$ is referred to as an
\emph{experience}, which is the basic unit of information describing a RL system. The control loop is then repeated
forever or terminated by reaching either a terminal state or a maximum time step $t=T$. The time frame from $t=0$ to the
terminal state (either a defined state/s or a maximum state) is called an \emph{episode}. In addition, the sequence of 
experiences over an episode, $\tau = (s_0, a_0, r_0), (s_1, a_1, r_1), \dots$ is known as a \emph{trajectory}. An agent 
typically needs many episodes to learn a good policy.

In a more formal way, we can describe states, actions and rewards as: (i) $s_t \in {\cal S}$ where ${\cal S}$ is the 
state space, (ii) $a_t \in {\cal A}$ where ${\cal A}$ is the action space, and (iii) $r_t = {\cal R}(s_t, a_t, s_{t+1})$ 
where ${\cal R}$ is the reward function. Here, the state space ${\cal S}$ is the set of all possible states in an 
environment. It can be defined as integers, real numbers, vectors, etc. Similarly, the action space ${\cal A}$ is the 
set of all possible actions. It is commonly defined as either a scalar or a vector. The reward function 
${\cal R}(s_t, a_t, s_{t+1})$, for its part, assigns a real number (positive or negative) to each transition. 
The state space, action space, and reward function are determined by the environment.

Let us now consider how an environment transitions from one state to the next using the \emph{transition
function}. In RL, a transition function is formulated as a Markov decision process (MDP), which means in practice that one
assumes that the transition to the next state $s_{t+1}$ only depends on the previous state $s_t$ and action $a_t$. This 
is known as the \emph{Markov property} and can be mathematically formulated as
\begin{equation} \label{eq-MDP}
	s_{t+1} \sim P(s_{t+1} | s_t, a_t) ,
\end{equation}
which means that the next state $s_{t+1}$ is sampled from a probability distribution $P(s_{t+1} | s_t, a_t)$.

With this new ingredient, we can now compile all MDP elements: 
${\cal S}$, ${\cal A}$, ${\cal R}(\cdot)$, $P(\cdot)$, where we recall ${\cal S}$ is the set of states, ${\cal A}$ is the 
set of actions, $P(s_{t+1} | s_t, a_t)$ is the transition function of the environment, and ${\cal R}(s_t, a_t, s_{t+1})$ 
is the reward function. Let us remark that RL algorithms tackled in this work are \emph{model-free}, this is, the agents have 
access to neither the transition function, $P(s_{t+1} | s_t, a_t)$, nor the reward function,
${\cal R}(s_t, a_t, s_{t+1}$). The only way in which an agent gets information about these functions is through the states,
actions, and rewards it actually experiences in the environment.

As previously indicated, to formulate a RL problem it is necessary to formalize the objective which the agent is intended 
to maximize. For this purpose, we first define the \emph{return} $G(\tau)$ using a trajectory from an episode, 
\begin{equation} \label{eq-return}
	G(\tau) = r_0 + \gamma r_1 + \gamma^2 r_2 + \dots + \gamma^T r_T = \sum^{T}_{t=0} \gamma^t r_t ,
\end{equation}
i.e., as a discounted sum of the rewards in a trajectory, where $\gamma \in [0,1]$ is the \emph{discount factor}. 
The discount factor is an important parameter which changes the way future rewards are considered. The smaller $\gamma$, 
the less weight is given to rewards in future time steps.

On the other hand, the \emph{objective} $J(\tau)$ is simply defined as the expectation of the returns over many trajectories 
evaluated with a given policy $\pi$, i.e.,
\begin{equation}
	J(\pi) = \mathbb{E}_{\tau \sim \pi} \left[G(\tau) \right] = \mathbb{E}_{\tau} 
	\left[ \sum^{T}_{t=0} \gamma^t r_t \right] .
\end{equation}
The expectation accounts for stochasticity in the actions and the environment.

A key question in RL concerns what an agent should learn. There are three basic properties that can be useful to an agent:
(i) a policy, (ii) a value function, and (iii) an environment model. First, if we recall, the policy $\pi$ is that which 
maps states to actions, which can be formalized with the notation $a \sim \pi(s)$. A policy can be stochastic and therefore, 
we can write this as $\pi (a | s)$ to denote the probability of an action $a$ given a state $s$. 

The \emph{value functions} provide information about the objective. They help an agent to understand how good the states and 
available actions are in terms of the expected future return, allowing to determine a policy from this information. 
There are two types of value functions defined as 
\begin{eqnarray}
	V^\pi(s) & = & \mathbb{E}_{s_{t^{\prime}}=s, \tau \sim \pi} \left[ \sum^T_{t=t^{\prime}} \gamma^t r_t \right] , 
    \label{eq-V} \\
	Q^\pi(s,a) & = & \mathbb{E}_{s_{t^{\prime}}=s, a_{t^{\prime}}=a, \tau \sim \pi} \left[ \sum^T_{t=t^{\prime}} 
    \gamma^t r_t \right] .\label{eq-Q} 
\end{eqnarray}
The \emph{state-value function} $V^\pi(s)$ in Eq.~(\ref{eq-V}) evaluates the quality of a state. It measures the expected 
return from being in state $s$, assuming the agent continues to act according to its current policy $\pi$. It is worth noting 
the return $G(\tau) =  \sum^T_{t=t^{\prime}} \gamma^t r_t$ is measured from the current state to the end of an episode. 
The \emph{action-value function} $Q^\pi(s,a)$ of Eq.~(\ref{eq-Q}) evaluates how good a state-action pair is. It measures 
the expected return from taking action $a$ in state $s$ assuming that the agent continues to act according to its current 
policy, $\pi$.

Finally, an environment model is summarized in the transition function $P(s_{t+1} | s_t,a_t)$ that provides information 
about the environment. If an agent learns this function, it is able to predict the next state $s_{t+1}$ that the environment 
will transition into after taking action $a$ in state $s$. Often, good models of the environment are not available and 
in this work we shall not make use of this type of function or the corresponding algorithms.

In RL, an agent learns a function of any of the previous properties to decide what actions to take with the goal of 
maximizing the objective. In most practical problems the different spaces -- state, action, etc. -- are so large that the 
key functions need to be approximated. Currently, the most popular methods to approximate these functions are based on 
deep neural networks, which gives rise to the concept of \emph{Deep Reinforcement Learning}. This is the method of choice 
in this work.

On the other hand, according to the three primary learnable functions in RL (see above), there are three major families 
of deep RL algorithms -- \emph{policy-based}, \emph{value-based}, and \emph{model-based methods} which learn policies, 
value functions, and models, respectively. In Section~\ref{sec-results}, we shall present the main results of this work 
organized according to the corresponding RL algorithm employed and we shall also include a brief description of the main 
characteristics of every used algorithm.

\section{Optimizing NFRHT between multilayer hyperbolic metamaterials} \label{sec-system}

\subsection{Physical problem}

In this Section we describe the specific problem that we have selected to illustrate the use of RL in the context of 
radiative heat transfer, namely the optimization of the near-field radiative heat transfer between multilayer hyperbolic
metamaterials~\cite{Garcia-Esteban2021Dec,Garcia-Esteban2024_GAN}. 

As discussed in the introduction, a major breakthrough in recent years in the field of thermal radiation has been the 
confirmation of the possibility to overcome Stefan-Boltzmann's law for the radiative heat transfer between two bodies by 
bringing them sufficiently close \cite{Polder1971Nov}. This physical phenomenon is due to the fact that in the near-field 
regime, bodies can exchange radiative heat via evanescent waves. This type of contribution is not considered in 
Stefan-Boltzmann's law and dominates the NFRHT for sufficiently small separations \cite{Song2015May,Cuevas2018Oct,Biehs2021Jun}. 
Different strategies have been recently proposed to further enhance NFRHT. One of the most prominent ones makes 
use of multiple surface modes that appear in multilayer structures where dielectric and metallic layers are alternated 
to give rise to the so-called \emph{hyperbolic metamaterials} \cite{Guo2012Sep,Biehs2012Sep,Guo2013Jun,Biehs2013Apr,Bright2014Jun,
Miller2014Apr,Biehs2017Feb,Iizuka2018Feb,Song2020Feb,Moncada-Villa2021Feb}. The hybridization of surface modes in different 
metal-dielectric interfaces can lead to a great enhancement of the NFRHT, as compared to the case of two infinite parallel
plates~\cite{Iizuka2018Feb}. 

\begin{figure}[h!]
\includegraphics[width=0.96\columnwidth,clip]{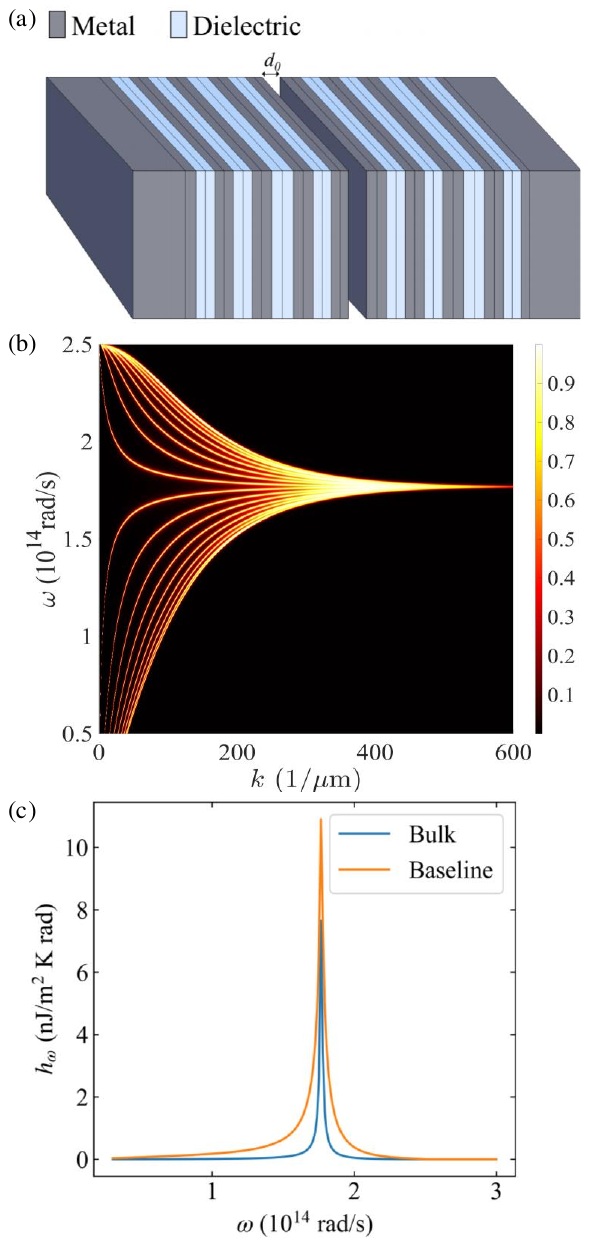}
\caption{(a) Schematic representation of the physical system under study. It features two identical hyperbolic 
metamaterials comprising alternating metallic (grey) and dielectric (blue) layers. Both reservoirs have infinitely-extended 
layers and are separated by a distance $d_0 = 10$~nm. Each layer has a thickness of 5~nm and both subsystems are backed 
by a metallic substrate. (b) Transmission of evanescent waves as a function of the frequency ($\omega$) and the 
parallel wavevector ($k$) for the periodic structure of panel (a) composed by 16 active layers per subsystem. 
(c) The corresponding spectral heat transfer coefficient $h_{\omega}$ at room temperature ($T=300$~K) as a function 
of the frequency, baseline in legend. The result is compared to that of two metallic plates (bulk) with the same gap.}
\label{fig-baseline}
\end{figure}

Following Ref.~\cite{Iizuka2018Feb}, here we consider the radiative heat transfer between two identical multilayer
structures separated by a gap $d_0$, see Fig.~\ref{fig-baseline}(a). Each thermal reservoir contains $N_l$ total layers 
alternating between a metallic layer with a permittivity $\epsilon_\mathrm{m}$ and a lossless dielectric layer of
permittivity $\epsilon_\mathrm{d}$. The thickness of the layer $i$ is denoted by $d_i$. The dielectric layers are 
set to vacuum ($\epsilon_\mathrm{d} =1$) and the metallic layers are described by a permittivity given by a Drude 
model: $\epsilon_\mathrm{m}(\omega) = \epsilon_{\infty} - \omega^2_p/[\omega (\omega + i \gamma)]$, where
$\epsilon_{\infty}$ is the permittivity at infinite frequency, $\omega_p$ is the plasma frequency, and $\gamma$ is
the damping rate. From now on, we set $\epsilon_{\infty} = 1$, $\omega_p = 2.5 \times 10^{14}$ rad/s, and 
$\gamma = 1 \times 10^{12}$ rad/s. With this choice of these parameters the surface plasmon frequency is similar 
to the surface phonon-polariton frequency of the interface between SiC and vacuum.

We describe the NFRHT between the hyperbolic metamaterials within the theory of fluctuational electrodynamics 
\cite{Rytov1953,Rytov1989}. In this system, the NFRHT is dominated by TM- or $p$-polarized evanescent waves and 
the heat transfer coefficient (HTC) between the two bodies, i.e., the linear radiative thermal conductance per 
unit of area, is given by \cite{Basu2009Oct}
\begin{equation}
h = \frac{\partial}{\partial T} \int^{\infty}_0 \frac{d\omega}{2\pi} \: \Theta(\omega, T) 
\int^{\infty}_{\omega/c} \frac{dk}{2\pi} \, k \: \tau_p(\omega, k) ,
\end{equation}
where $T$ is temperature, $\Theta(\omega, T)= \hbar \omega/ (e^{\hbar \omega/ k_{\rm B} T} -1)$ is the mean thermal 
energy of a mode of frequency $\omega$, $k$ is the magnitude of the wave vector parallel to the surface planes, and
$\tau_p(\omega, k)$ is the transmission (between 0 and 1) of the $p$-polarized evanescent modes given by
\begin{equation}
 \tau_p(\omega, k) = \frac{4 \left[ \mbox{Im} \left\{ r_p(\omega,k) \right\} \right]^2 e^{-2q_0 d_0}}
 {| 1 - r_p(\omega,k)^2 e^{-2 q_0 d_0} |^2} .
\end{equation}
Here, $r_p(\omega,k)$ is the Fresnel reflection coefficient of the $p$-polarized evanescent waves from the vacuum 
to one of the bodies and $q_0 = \sqrt{k^2 - \omega^2/c^2}$ ($\omega/c < k$) is the wave vector component normal to 
the layers, in vacuum. The Fresnel coefficient needs to be computed numerically and we have done it by using the 
scattering matrix method described in Ref.~\cite{Caballero2012Jun}. In our numerical calculations of the HTC we also 
took into account the contribution of $s$-polarized modes, but it turns out to be negligible for the gap sizes explored 
in this work.

The interest in the NFRHT in these multilayer structures resides in the fact that the heat exchange in this regime is
dominated by surfaces modes that can be tuned by playing with the layer thicknesses. In the case of two parallel 
plates made of a Drude metal, the NFRHT is dominated by the two cavity surface modes resulting from the hybridization 
of the surface plasmon polaritons (SPPs) of the two metal-vacuum interfaces \cite{Iizuka2018Feb}. These two cavity 
modes give rise to two near-unity lines in the transmission function $\tau_p(\omega, k)$. If we introduce more internal 
layers, we can have additional NFRHT contributions from surface states at multiple surfaces. This is illustrated in 
Fig.~\ref{fig-baseline}(a) for the case of $N = 16$ active layers with $d_i = 5$ nm and a gap size $d_0 = 10$ nm. Apart from
the active layers, both reservoirs contain an additional 5 nm-thick metallic layer in the outer part to properly define the 
gap, as well as a semi-infinite metallic substrate on the other side. This example, in which we have in practice a periodic structure 
with 8 physical layers (4 metallic and 4 dielectric layers) with thickness $d_i = 10$ nm, exhibits multiple near-unity resonances 
in the transmission function $\tau_p(\omega, k)$, see Fig.~\ref{fig-baseline}(b). These contributions resulting from 
additional surface states originating from internal layers lead to a great enhancement of the NFRHT as compared 
to the bulk system (two parallel metallic plates) in a wide range of gap values \cite{Iizuka2018Feb}. This is illustrated 
in Fig.~\ref{fig-baseline}(c) where we show the spectral HTC, $h_\omega$, defined as the HTC per unit of frequency: 
$h = \int^{\infty}_0 h_\omega \: d\omega$, for both the system pictured in Fig.~\ref{fig-baseline}(a) (labeled as 
baseline) and the bulk system with the same gap $d_0 = 10$~nm.

Our concrete goal is to maximize the HTC between these two hyperbolic metasurfaces by finding the optimal configuration
of alternating dielectric and metallic layers (number and thickness). We keep fixed the gap size to $d_0 = 10$~nm, the total 
active thickness of the multilayer areas, and assume room temperature ($T=300$ K). We also assume the two multilayer systems 
to be identical since any asymmetry tends to reduce the HTC.

\subsection{RL formulation of the optimization problem} \label{sec-RL formulation}

We now describe how we tackle our optimization problem in the spirit of RL, which requires to formulate it as a sequential 
decision-making problem. The physical system is composed of two identical layered structures which, unless we state otherwise, 
contain 16 active layers with a thickness of 5 nm. The two subsystems are separated by a gap $d_0 = 10$ nm, each having a 
semi-infinite metallic substrate. 

Irrespective of the employed algorithm, we define the central RL concepts as follows:
\begin{enumerate}
    \item \textbf{Goal}: our goal is to maximize the HTC via the modification of the layer configuration.
    
    \item \textbf{State}: each state describes a layer configuration. We define the material by an integer label, 
    0 for the dielectric and 1 for the metal. Thus, a state is represented by a vector of 0s and 1s with 16 components, 
    each representing one of the 5 nm-thick layers.  
    
    \item \textbf{Action}: the action space is an ensemble of two decisions made concurrently, namely which layer to 
    study and what material to consider for it. This includes the possibility for the configuration to remain unchanged.
    
   \item \textbf{Reward}: the reward is the HTC corresponding to the next state in units of $10^5$ W/m$^2$K. Thus, 
   for this problem, ${\cal R} = {\cal R}(s_{t+1})$ only. For better performance, we consider as a baseline for the 
   reward values the HTC of our physically intuitive ``best guess" configuration, which corresponds to the periodic photonic 
   crystal shown in Fig.~\ref{fig-baseline}(a). Any positive reward implies that we have found a higher HTC value.
   
   \item \textbf{Episode Termination}: we impose an episodic formulation by defining a fixed number of actions taken 
   in a trajectory before resetting to the initial state, containing all 0s (all dielectric layers except the 5 nm-thick metallic 
   layer defining the gap). This enables the network to perform much more training on known states, and to finish the optimization 
   in an acceptable number of steps. We take the length of an episode to be twice or four times the number of layers of an state, 
   so any existing state is potentially reached comfortably.
\end{enumerate}

\section{Results} \label{sec-results}

In this section we describe the main results obtained for the optimization of the NFRHT between the multilayer
hyperbolic structures described in the previous section. For didactic reasons, we present separately the 
results obtained with the different RL algorithms and in every subsection we describe the basics of the corresponding 
method alongside with a discussion of the peculiarities concerning their application to our problem.  

\subsection{Value-based algorithms: SARSA, deep Q-learning and extensions} \label{sec-value-based}

As value-based algorithms are historically the most widely used and discussed in RL~\cite{SuttonBarto}, we shall address 
first their formulation and usage. Value-based algorithms are based on two core ideas. The first one is \emph{temporal 
difference} (TD) learning, which is an alternative to the use of Monte Carlo sampling for gathering experiences from an
environment (see Sec.~\ref{sec-REINFORCE}) to estimate state/state-action values. The key idea in TD learning is that 
state/state-action values are defined recursively, that is, their value in a given time step is defined in terms of the value 
in the next time step. This makes TD learning an useful method for backing up the reward information from later to earlier 
steps through time. As state/action-value functions represent an expectation over different trajectories, this leads to the
display of a lower variance than Monte Carlo sampling.

The second idea has to do with the famous \emph{exploration-exploitation trade-off} in RL. When the agent is learning an 
estimate of the state/state-action values, the usage of this estimate can lead to better returns (exploitation). However, 
if one always selects actions based on current values, which might be far from the optimal ones, this would lead to a 
deterministic behavior that can prevent the agent from discovering better unknown actions (exploration). This 
exploration-exploitation trade-off is a key challenge in RL and it can be addressed employing \textit{stochastic policies}, 
where the exploration can be distributed along all the training and, as the estimation gets better, it gradually shifts 
closer to a deterministic policy. An example of an stochastic policy, which will be used in this work, is the 
$\varepsilon$-greedy policy, where the agent explores with a probability of $\varepsilon$ and exploits with a probability 
of $1-\varepsilon$.

\subsubsection{SARSA}

\emph{SARSA} is one of the oldest RL value-based algorithms and, despite its limitations (see below), it is convenient to start
by describing its use for our problem. This algorithm is based on the estimation of the action-value function or $Q$-function. 
It employs TD learning to produce the target state-action values or $Q$-values, from now on denoted as $Q_{\rm tar}$. 
Therefore, it combines the reward given by the environment, $r_{t}$, with the $Q$-value estimates of the next state which 
approximate the remaining part of the expected return. This is summarized in the following update rule:
\begin{equation} \label{eq:Sarsa-update}
Q_{\rm tar}^{\pi}(s_t, a_t) = r_{t} + \gamma Q^{\pi}(s_{t+1}, a_{t+1}) .
\end{equation}
Notice that the $Q$-value estimate depends on the following action, as we base our estimates solely on state-action estimates. 
Over the course of many examples, the proportion of selected actions given an state will approximate the probability 
distribution over all actions.

In practice, we employ a neural network for the approximation of the $Q$-function, the \textit{$Q$-network}, which returns the 
$Q$-value estimates of the selected state-action pairs. As a consequence, each update of the $Q$-value is not complete, as 
neural networks learn gradually using gradient descent, moving partially towards the target value. With all that, SARSA algorithm 
is summarized in pseudocode~\ref{pseudocode-Sarsa}.

\begin{algorithm}[b] \label{pseudocode-Sarsa} 
\vspace{2mm}
    \caption{SARSA pseudocode \cite{SuttonBarto, DeepRLbook}.}
    \KwIn{a differentiable action-value function parametrization $\widehat{q} : S \times A \times \mathbb{R}^d 
    \rightarrow \mathbb{R}$}
    \KwIn{algorithm parameters. 
    Initialize learning rate $\alpha > 0$, epsilon $1 \geq \varepsilon > 0$, discount rate $1 \geq \gamma \geq 0$.}
    \KwOut{optimized $\widehat{q}$}
    Initialize arbitrarily $\widehat{q}$ weights, $\vec{w} \in \mathbb{R}^d$
    
    Initialize $s_0 \neq terminal$
    
    \For {training step}{
        Generate a new batch of episodic experiences $s_t$, $a_t$, $r_{t}$, $s_{t+1}$, $a_{t+1}$ with 
        $\varepsilon$-greedy policy wrt. 
        $\widehat{q}(s_t, \cdot)$
        
        \For {experience in batch}{
            Store $estimation(\vec{w})$: $\widehat{q}(s_t, a_t)$   
            
            Store $target$: $r_{t} + \gamma \widehat{q}(s_{t+1}, a_{t+1})$
        }
        $\vec{w} \leftarrow \vec{w} + \alpha \nabla_{\vec{w}} $loss($estimation(\vec{w}), target$)  
        
        Decay $\varepsilon$  
        
        Decay \textit{learning rate}
    }
\end{algorithm}

The workflow of SARSA is similar to a supervised learning workflow, in which each estimate has a target value to reach and, 
with it, we can evaluate how well our neural network is performing and thus reduce the discrepancies between the values. 
In this sense, we use an iterative approach to improve the $Q$-value, as we can explicitly see in line 3. Notice from lines 
5 and 7 that, as SARSA is based on TD learning, only the information from the next step is required to form the target of 
the current $Q$-value, allowing to update the $Q$-function in a batched manner. Regarding sample efficiency, we can see
from line 4 that the next action $a_{t+1}$ is obtained with the same policy used to gather the previous action $a_t$, this is, 
$\varepsilon$-greedy policy over the only $Q$-network of the algorithm. This specific selection of the next action makes 
SARSA an \textit{on-policy} algorithm, that is, an algorithm in which the information for improving the policy ($a_{t+1}$)
depends on the policy to gather data ($a_{t}$). Because this on-policy behavior, each training iteration can only 
use experiences obtained following the current policy, so each time the $Q$-network parameters are updated, all experiences 
must be discarded and new experiences have to be collected, as reflected again with line 4 and its position within the 
training loop. Finally, something that arises from the use of currently collected experiences for the estimation of the target 
$Q$-value is the high correlation between experiences, as the data used to update the network is often from a single 
episode, which can lead to high variance in different parameter updates.

\begin{figure}[t]
    \includegraphics[width=\columnwidth]{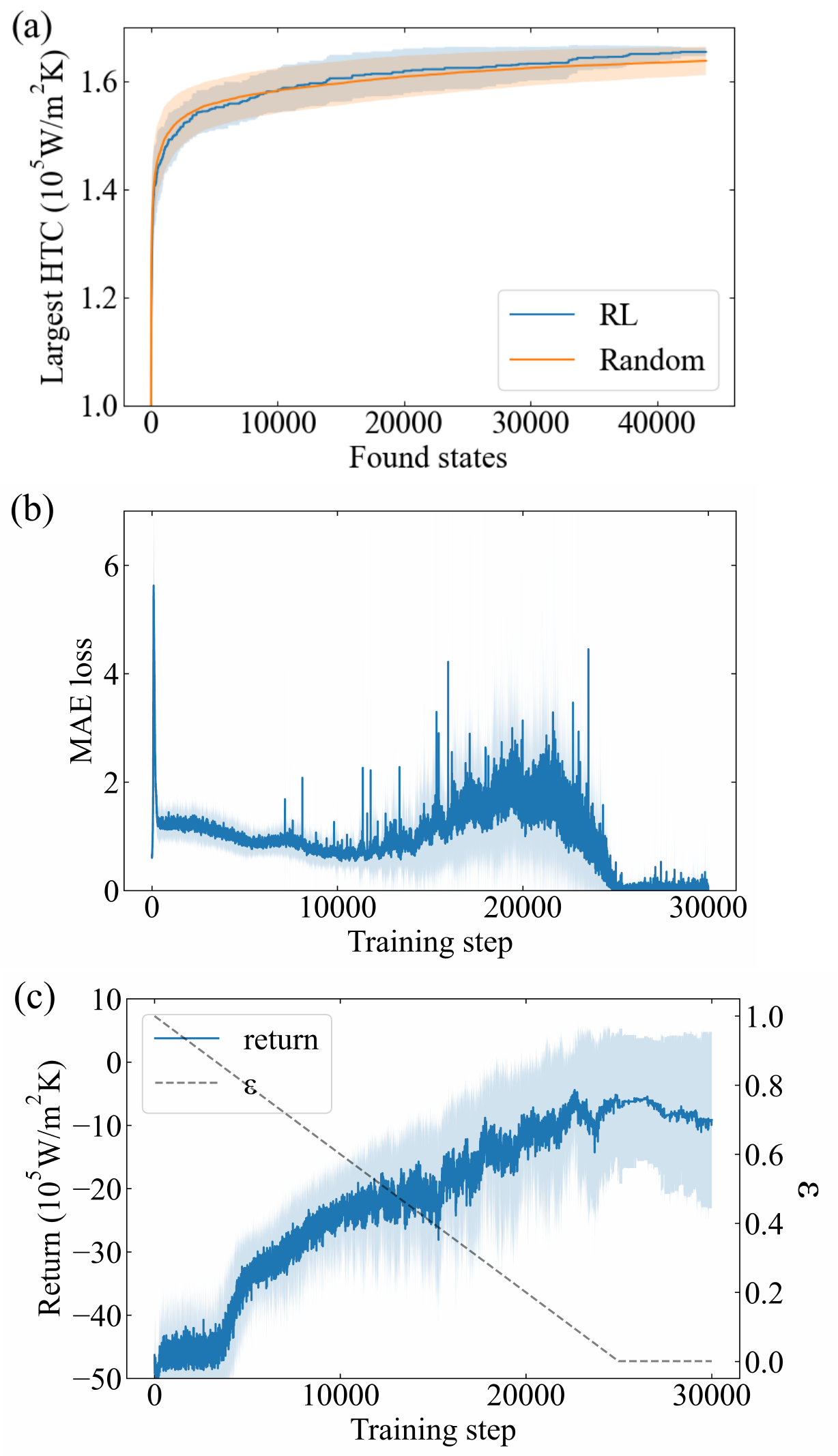}
    \caption{Training of SARSA algorithm for our physical problem of interest. (a) Largest HTC discovered as a function 
    of the number of found states in the problem with 16 layers obtained with SARSA algorithm. We also present the results 
    obtained with the random algorithm. (b) The evolution of the corresponding loss curve of SARSA algorithm. 
    (c) Return obtained in a simulation of an episode with the $Q$-network of SARSA algorithm at each training step.
    The dashed line corresponds to the value of $\varepsilon$ (right scale). In all panels the solid lines  correspond to 
    the mean value and the shaded areas to the standard deviations, as obtained in 40 independent runs for SARSA algorithm.}
    \label{fig-sarsa}
\end{figure}

In Fig.~\ref{fig-sarsa} we present a summary of the results obtained with SARSA for our hyperbolic multilayer system with 
16 layers, which includes 40 independent runs, represented with their mean and standard deviation. These results were 
obtained with the hyperparameters specified in Table~\ref{tab-hyper-SARSA} of Appendix~\ref{sec-hyperpara}. 
Figure~\ref{fig-sarsa}(a) shows the largest HTC obtained as a function of the number of found states. To gauge the quality of 
our method, we compare SARSA results in this panel with an algorithm in which different states are randomly selected and the 
maximum HTC is recorded as the algorithm progresses. This random algorithm is particularly efficient for relatively small state 
spaces, as it is forced to always find new states. Therefore, its results constitute a good test for the different RL algorithms. 
To ensure reliable statistics, its mean value and deviation were obtained with 1000 independent runs in all cases.

In Fig.~\ref{fig-sarsa}(b) we present the evolution of the loss function of the $Q$-network along the training proceeds.
Finally, Fig.~\ref{fig-sarsa}(c) displays the return in a greedy simulation of an episode with the $Q$-function obtained 
each training step. Recalling the basis of RL (see Sec.~\ref{sec-RL}), the usual goal of RL is to maximize the expected return $G(\tau)$. 
In this sense, we can see in Fig.~\ref{fig-sarsa}(c) that the return increases along the exploratory phase, so SARSA algorithm is achieving
better policies as the training proceeds, as expected. Regarding the loss function, Fig.~\ref{fig-sarsa}(b) shows that its general
tendency is indeed of decrease, indicating that the $Q$-network tends to converge, as desired. However, some noise during the 
training can be noticed. When $\varepsilon$ is close to 0.5 the average loss starts increasing, and when $\varepsilon$ reaches its minimum 
value, the loss rapidly adopts its lowest value. Apart from the variance arising just from SARSA formulation due to the high
correlation between the experiences of each batch, the noise seems to arise from the greedy behavior of the method in 
intersection with the shape of the state-action space. In our case, the noise could indicate the $Q$-value of the state-actions 
visited through the greedy policy is not that close to the value of the previously seen state-action pairs. With that, an overfitting of the 
$Q$-network is thought to be made to the state-actions of the greedy behavior, which increases the mean loss when those states 
are not that frequently visited and gradually decreases the loss as the greedy behavior is more prominent.

On the other hand, Fig.~\ref{fig-sarsa}(a) shows that SARSA is capable of finding the optimal configuration for our problem, 
but it performs similarly to the random algorithm with the exception of the end of the curve, where SARSA runs exhibit a smaller
variance. Thus, the main conclusion from this analysis is that although SARSA can learn an improved policy, it is not sample efficient 
enough to achieve our objective for the selected environment. It is important to emphasize that, as in any machine learning 
problem, we cannot rule out that with a better selection of hyperparameters SARSA could clearly beat the random algorithm, 
especially in problems of higher dimensionality. In any case, as we shall see in the next subsection, sample efficiency can be 
notably improved using other types of value-based algorithms, so we shall not dwell too much here with SARSA algorithm.

\subsubsection{Q-Learning}

As in SARSA algorithm, \emph{$Q$-learning} is based on TD learning in order to obtain the target value. In this case, the 
update rule for the $Q$-values reads
\begin{equation} \label{eq-Q-learning update}
Q_{\rm tar}^{\pi}(s, a) = r_{t} + \gamma \max_{a_{t+1}} Q^{\pi}(s_{t+1}, a_{t+1}) .
\end{equation}
Notice that the selection of the following action $a_{t+1}$ is made with a max operator, which indicates that we are taking the 
action that maximizes the $Q$-value of the next state. This might be seen as a small change with respect SARSA, but it has important 
consequences, overcoming some of SARSA limitations. With it, in Eq.~(\ref{eq-Q-learning update}) we are learning the optimal 
$Q$-function instead of the $Q$-function of the current policy as in SARSA, improving so the stability and speed of learning. 
In addition, this makes $Q$-learning an \textit{off-policy} algorithm, as the information used to learn the $Q$-value is 
independent of the policy used for gathering data. Therefore, off-policy behavior allows us to learn from experiences gathered 
by any policy. It allows to reuse and decorrelate experiences, reducing the variance of each update and improving the sample
efficiency with respect to SARSA.

It is interesting to underline that, for our application, the reuse of experiences is of great importance as our goal is to
obtain the maximum HTC with as few explored states as possible. In addition, the usage of neural networks makes this aspect 
even more relevant as they rely on gradient descent, for which each parameter update must be small because the gradient
only returns meaningful information near employed parameters. This makes the possibility of reusing experiences 
important as the network's parameters may need to be updated multiple times.

\begin{algorithm}[b] \label{pseudocode-Q-learning} 
    \caption{$Q$-Learning pseudocode \cite{SuttonBarto, DeepRLbook}.}
    \vspace{2mm}
    \KwIn{a differentiable action-value function parametrization $\widehat{q} : S \times A \times \mathbb{R}^d \rightarrow 
    \mathbb{R}$}
    \KwIn{algorithm parameters.
    Initialize learning rate $\alpha > 0$, epsilon $1 \geq \varepsilon > 0$, discount rate $1 \geq \gamma \geq 0$.}    
    \KwOut{optimized $\widehat{q}$}
    Initialize arbitrarily $\widehat{q}$ weights, $\vec{w} \in \mathbb{R}^d$ 
    
    Initialize $s_0 \neq terminal$
    
    \For {training step}{
        Generate a batch of episodic experiences $s_t$, $a_t$, $r_{t}$, $s_{t+1}$ with $\varepsilon-greedy$ behavior 
        policy wrt. 
        $\widehat{q}(s_t, \cdot)$
        
        \For {experience in batch}{
            Store $estimation(\vec{w})$: $\widehat{q}(s_t, a_t)$
            
            Choose $a_{t+1}$  with $greedy$ target policy wrt. $\widehat{q}(s_{t+1}, \cdot)$   
            
            Store $target$: $r_{t} + \gamma \widehat{q}(s_{t+1}, a_{t+1})$
            }
        $\vec{w} \leftarrow \vec{w} + \alpha \nabla_{\vec{w}} $loss($estimation(\vec{w}), target$)
        
        Decay $\varepsilon$
        
        Decay \textit{learning rate}
    }
\end{algorithm}

We summarize the $Q$-learning algorithm in pseudocode~\ref{pseudocode-Q-learning}. Comparing lines 4 and 7, one sees that
$Q$-learning gathers experiences selecting the action $a_t$ with an $\varepsilon$-greedy behavior, while it estimates the target 
$Q$-value with a greedy selection of $a_{t+1}$, as previously seen in update rule of Eq.~(\ref{eq-Q-learning update}). Although
those lines are the only difference with SARSA, they enable to further implement all previously mentioned advantages like 
experience reuse with the extensions described below. We shall present in this work only the $Q$-learning results obtained 
with the most refined version considered of $Q$-learning algorithm. Therefore, we shall postpone the discussion of the $Q$-learning 
results and introduce in what follows several sophistications for the naive version of this algorithm.

\subsubsection{Q-Learning extensions}

As mentioned above, $Q$-learning is potentially better than SARSA to achieve our goal because of its off-policy nature \cite{SuttonBarto}. 
Building upon this nature, we shall make use here of several modifications of $Q$-learning that have been proposed to 
enhance its sample efficiency and stability:

\textit{Experience replay.-} Introduced by Lin~\cite{lin1992self}, this idea consists of storing experiences in a 
    memory in order to reuse them even if they were taken with old policies, allowing for more efficient learning from a reduced 
    number of gathered experiences. In practice, a \textit{experience replay memory} stores the agent's most recent experiences 
    up to a given memory size, large enough to contain many episodes, replacing the oldest experiences by the newest ones once 
    this size is reached. With that, every time an agent needs batches to be trained, it retrieves them from replay memory in a 
    random-uniformly manner. Then, each one of the batches is used to update the training network. Like this, in addition to 
    introduce higher sample efficiency, we ensure we have decorrelated experiences for training as they are likely to be from 
    different policies and episodes, contrary to SARSA, stabilizing the training as we reduce the variance of parameter updates.
    Finally, to set a widely used notation, the combination of Q-learning with the usage of Q-networks and the presented memory
    replay technique receives the name of \emph{Deep Q-Networks} (DQN), set by Mnih et al~\cite{mnih2013atari}.  We use this
    notation along the rest of the paper.

\textit{Target network.-} Introduced by Mnih \emph{et al.}~\cite{mnih2015human}, it focuses on reducing the changes in 
    the target value by means of a second network, called the \textit{target network}. The idea is to use a lagged copy of the 
    training network which update frequency is less than that of the training network. Then, this secondary network is used to 
    generate the state-action estimate for the target value, $\max_{a_{t+1}} Q^{\pi}(s_{t+1}, a_{t+1})$, stopping its value from 
    moving. This idea addresses the issue where the target is constantly changing because of network updates, stabilizing the 
    training and making divergences less likely as it avoids ambiguity regarding the values the network must approach.
    
\textit{Double Q-learning.-} Introduced by van Hasselt \emph{et al.}~\cite{vanhasselt2015,vanhasselt2010}, the basic 
    idea is using two different networks trained with different experiences for the estimation of the next $Q$-value used for 
    obtaining the target value. This double estimation is computed using a network for retrieving the maximizing action, and the 
    remaining network for producing the $Q$-value with the selected action, as follows
    \begin{equation} \label{eq-double Q-learning update}
    Q_{\rm tar}^{\pi}(s, a) = r_{t} + \gamma Q_{2}^{\pi}(s_{t+1}, \argmax_{a_{t+1}} Q_{1}^{\pi}(s_{t+1}, a_{t+1})).
    \end{equation}
    It mitigates the systematic overestimation of the state-action values by the deep $Q$-learning algorithm. This effect 
    arises from the use of an approximated algorithm. As it does not return a perfect estimation, if $Q$ estimations contain 
    any errors, maximum state-actions are likely to be positively biased, resulting in an overestimation of the $Q$-values as 
    Hasselt \emph{et al.}\ showed in their paper~\cite{vanhasselt2015}. With that, if we introduce the usage of a second 
    network trained with different experiences, we can remove the positive bias in the estimation. As with the introduction 
    of the target network we already have a second network and we want to avoid sampling more experiences, a common practice 
    is the usage of it as the secondary network. Although it is just a lagged copy of the training network, if the update 
    frequency of the target network is sufficiently low, it is considered to be different enough from the training network 
    to function as a different one.

\begin{algorithm}[t] \label{pseudocode-DQ-learning} 
\vspace{2mm}
    \caption{Double $Q$-learning with memory replay and target network pseudocode \cite{SuttonBarto,DeepRLbook}.}
    \KwIn{differentiable action-value function parametrizations $\widehat{q}, \tilde{q} : S \times A \times \mathbb{R}^d \rightarrow \mathbb{R}$}
    \KwIn{algorithm parameters.
    Initialize learning rate $\alpha > 0$, epsilon $1 \geq \varepsilon > 0$, discount rate $1 \geq \gamma \geq 0$, 
    new experiences per episode $h > 0$, batches per training step $B > 0$, 
    target network update frequency $F > 0$.}
    \KwOut{optimized $\widehat{q}$}
    Initialize arbitrarily $\widehat{q}$ weights, $\vec{w} \in \mathbb{R}^d$ 
    
    Equal target network $\tilde{q}$ weights to $\widehat{q}$ weights,  $\vec{\varphi} \in \mathbb{R}^d = \vec{w} \in \mathbb{R}^d$
    
    Initialize $s_0 \neq terminal$
    
    Initialize memory replay with a batch of episodic experiences $s_t, a_t, r_{t}, s_{t+1}$ with $\varepsilon$-greedy 
    behavior policy wrt. $\widehat{q}(s_t, \cdot)$
    
    \For {training step}{
        Store in memory $h$ episodic experiences $s_t$, $a_t$, $r_{t}$, $s_{t+1}$ with $\varepsilon-greedy$ behavior policy wrt. 
        $\widehat{q}(s_t, \cdot)$
        
        \For {batch B}{
            Sample a batch of experiences from memory
            
            \For {experience in batch}{
                Store $estimation(\vec{w})$: $\widehat{q}(s_t, a_t)$
                
                Choose $a_{t+1}$  with $greedy$ target policy wrt. $\widehat{q}(s_{t+1}, \cdot)$     
                
                Store $target$: $r_{t} + \gamma \tilde{q}(s_{t+1},a_{t+1})$
                }
            $\vec{w} \leftarrow \vec{w} + \alpha \nabla_{\vec{w}} $loss($estimation(\vec{w}), target$)
        }
        \If {training step $\propto$ F frequency}{
            Update target network $\tilde{q}$ weights, $\vec{\varphi} = \vec{w}$
        }
        Decay $\varepsilon$
        
        Decay \textit{learning rate}
    }
\end{algorithm}

Taking into consideration all these modifications, the final Double DQN algorithm we use for our application is summarized 
in the pseudocode~\ref{pseudocode-DQ-learning}. Let us emphasize the main differences with respect to the $Q$-learning
algorithm in pseudocode~\ref{pseudocode-Q-learning}. Lines 4, 6 and 8 describe the usage of memory replay, first initializing
it to gradually add more experiences at each training step. For training, $B$ batches of experiences are sampled, further 
leveraging experiences reuse compared to not using several batches. Next, lines 2, 12 and 15 refer to the usage of a target 
network, which is updated to the weights of the training network with a frequency $F$. Finally, double estimation is 
reflected in lines 11 and 12, where $a_{t+1}$ is taken with the training network $\widehat{q}$, while the next state-action 
value is obtained through the target network $\tilde{q}$.

\begin{figure}[t!]
    \includegraphics[width=0.95\columnwidth]{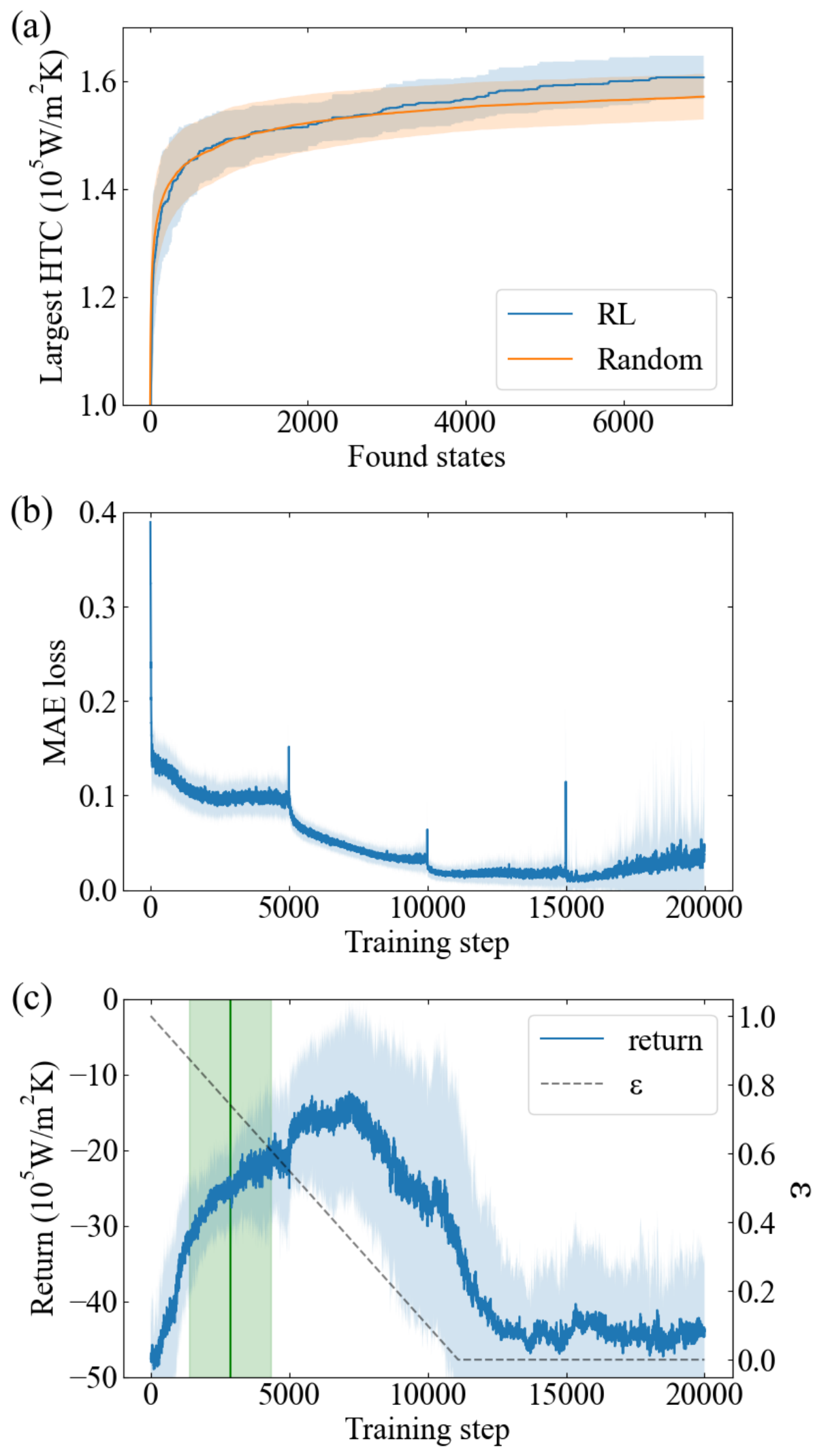}
    \caption{Training of the Double DQN algorithm. (a) Largest HTC discovered as a function of the number of found states in 
    the problem with 16 layers obtained with the Double DQN algorithm. We also present the results obtained with the random 
    algorithm. (b) Evolution of the corresponding loss curve of the Double DQN algorithm. 
    (c) Return obtained in a simulation of an episode with the $Q$-network of Double DQN algorithm at each training step.
    The dashed line corresponds to the value of $\varepsilon$ (right scale). The vertical line and the green shaded area 
    correspond to % the mean value and standard deviation of 
    the training steps at which the highest HTC of the runs are found.
    In all panels the solid lines  correspond to the mean value and the shaded areas to the standard deviations, as 
    obtained in 40 independent runs. In all cases, 4 experiences were stored per training step.}
    \label{fig-double DQN}
\end{figure}

Making use of this final \emph{Double DQN algorithm} and the hyperparameters listed in Table~\ref{tab-hyper-doubleDQN}, we 
obtained the results summarized in Fig.~\ref{fig-double DQN} for our multilayer system, again with 40 independent runs for 
Double DQN algorithm. In Fig.~\ref{fig-double DQN}(a) we display the maximum HTC for Double DQN and the random algorithm as 
a function of the number of found states. The Double DQN algorithm surpasses the random algorithm for a relatively small amount 
of explored states, $\sim 7000$  versus the $\sim 65000$ possible number of states of the system. To emphasize the quality of 
these results, let us say that for the 40 runs of Double DQN algorithm, the top 5 best HTC values for this system are found 
in $77.5\%$ of the runs and the best possible state in $35.0\%$ of the runs, while for the 1000 runs of the random algorithm 
these values are found in $44.3\%$ and $12.9\%$ of the runs, respectively.

\begin{figure}[t]
    \includegraphics[width=0.95\columnwidth]{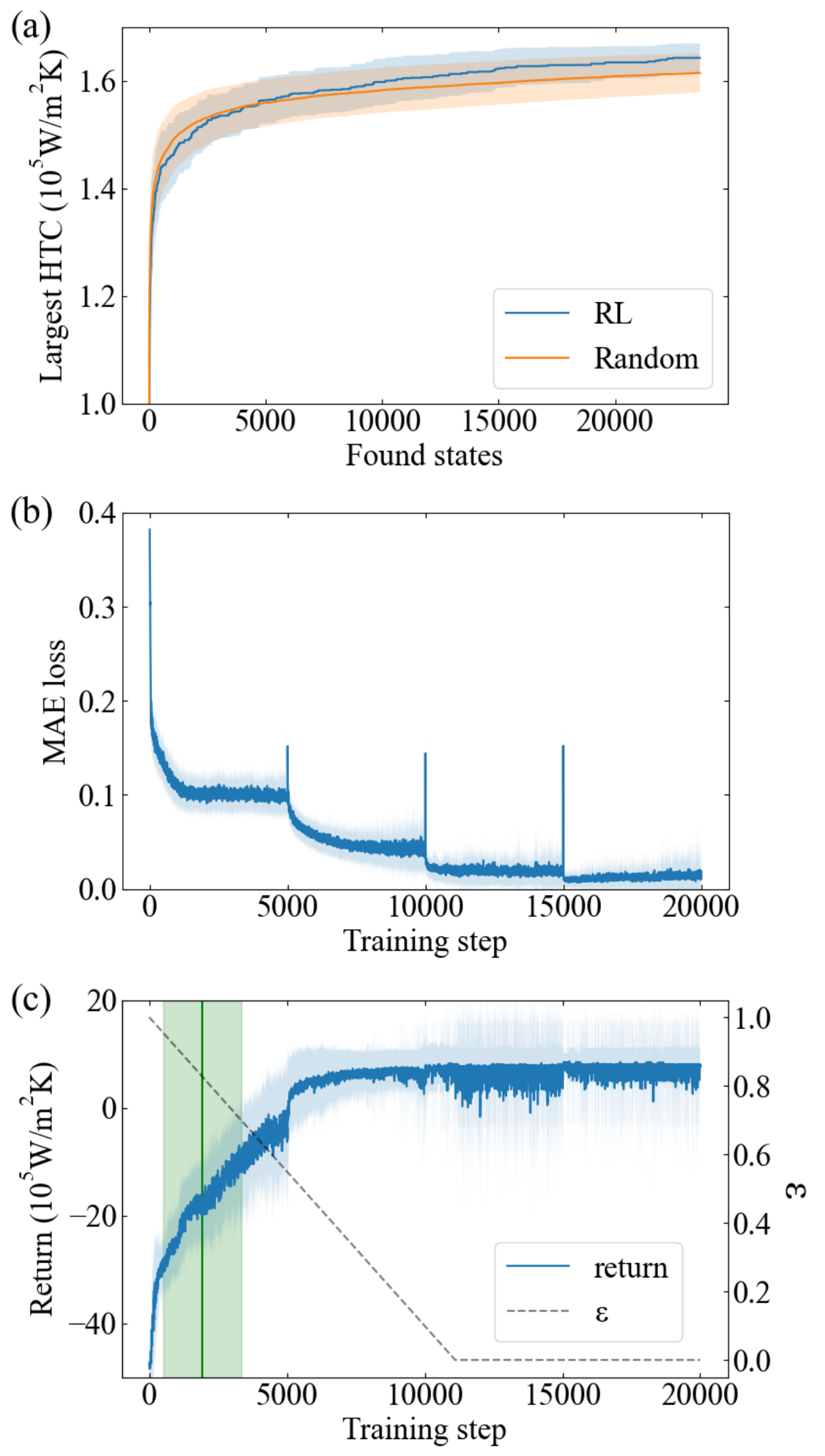}
    \caption{Same as in Fig.~\ref{fig-double DQN} but with 25 experiences stored per training step.}
    \label{fig-double DQN trained}
\end{figure}

To gain some insight into the training process of our algorithm, we display in Fig.~\ref{fig-double DQN}(b) the 
evolution of the loss of the $Q$-network. This loss decreases almost monotonically, which indicates that our algorithm 
is training. However, some irregularities can be appreciated again. First, we can see some regular peaks. These 
just correspond to the update of the target network, which produces the sudden change of the target. In addition, some 
noise appears at the end of the curve, which we believe it could arise from two effects. First, from the same effect as in 
SARSA's loss, i.e., the overfitting of the $Q$-network. Second, it could be due to the target network is not being that frequently
updated, leading to a target with less information. Therefore, we can end again with a less generic $Q$-network, 
which can lead to higher values of the loss.

Finally, as previously discussed, an important metric for the performance of the algorithm is the return 
in a greedy simulation with the $Q$-function obtained each train step, see Fig.~\ref{fig-double DQN}(c). Notice that  
the return first increases but ends up decreasing, which could appear to be upsetting. Let us recall that the return is 
the main result for RL applications with the standard algorithm's objective: to obtain a policy that maximizes the return 
of the system. However, our final objective here is slightly different: to explore the optimal state with as few explored states 
as possible. This is why the observed decay in the return is not worrisome in our case, although it means that we end up 
having a non-optimal policy.

In Fig.~\ref{fig-double DQN}(c) we also show as a vertical green line with a shaded area both the mean and standard 
deviation of the training step at which the highest HTC is discovered in the different runs. Two things are 
worth remarking: in this region $\varepsilon$ still has a sizeable value, so the algorithm still has chances of discovering
better states, and the return is still growing and is higher than the return of a completely exploratory policy, so we have 
a policy with some learning. This supports the fact that the decrease of the return is something not to worry about: with 
selected hyperparameters, our algorithm uses the policy learned at early stages to explore better states.

Although the decay of the return is not something critical in our case, it is important to understand why it occurs. A 
possible explanation is the no convergence to the optimal $Q$-function. As mentioned in Sec.~\ref{sec-RL formulation}, our reward 
is the difference of the HTC of the periodic multilayer system and that of the next state. In Fig.~\ref{fig-double DQN}(c), 
we can see that the return does not surpass the zero value, so we are not close to the optimal policy and, therefore, to 
the optimal $Q$-function. The reason why a below zero value is not close to a simulation of an episode of the optimal policy is 
the following. Knowing the optimal state (see Sec.~\ref{sec-conclusions}) and given a large episode length of 64 steps, an 
intuitively good policy would imply transitioning with as few number of steps as possible from the start state to the 
optimal one. It would result into transitioning to the optimal state in just 7 steps. Therefore, this policy would output 
7 unknown rewards and a reward of $0.28 \times 10^5$ W/m$^2$K during 57 steps. In addition to the best reward, we know the 
worst possible one, which has a value of $-1.37 \times 10^5$ W/m$^2$K. With it, the 7 unknown rewards must be equal or 
higher to the worst possible reward. Therefore, the good policy we have imagined has a return $G(\tau) \geq 6.37 \times 10^5$ 
W/m$^2$K, which has a positive value. By definition, the optimal policy is such that its expected return is greater or equal 
than any of the remaining existent policies for all states~\cite{SuttonBarto}. With this example, we have found a policy 
whose expected return from the initial state is higher than zero, which is over the return displayed at Fig.~\ref{fig-double DQN}(c). 
Thus, this demonstrates that we have not reached the optimal policy in that figure.

As we have not reached the optimal policy and therefore the true value of the $Q$-function, we only have an imperfect estimation 
of it. This estimation can help us reach the optimal policy during training. However, if we exploit 
it instead of using it to continue looking for the optimal policy, we can end up overtraining our network with experiences 
that a good policy is not likely to visit. This puts our policy farther from the optimal $Q$-function, losing the part of 
the estimation that was towards the good policy and, finally, turning the policy into a worst one because we are following 
non-optimal state-actions.

An interesting question at this stage concerns the issue of the impact of having a better estimation of the $Q$-values with 
the usage of more experiences and exploit its knowledge to obtain the optimal state. To elucidate this issue, we present in 
Fig.~\ref{fig-double DQN trained} the results obtained for the Double DQN algorithm using now 25 experiences stored in the 
memory replay per training step, rather than 4 as presented in Fig.~\ref{fig-double DQN}. In Fig.~\ref{fig-double DQN trained}(b), 
the loss function of the neural network still decays as expected with already known irregularities, while in 
Fig.~\ref{fig-double DQN trained}(c), the return increases up to a maximum value and stays there, as desired in regular 
applications of RL. This can mean that a sufficiently good policy is reached, so following it does not put us farther from 
the state-actions an optimal or suboptimal policy would follow. However, something must be noticed from Fig.~\ref{fig-double DQN trained}(a), 
namely higher HTC values with respect the random algorithm are now discovered when more states have been explored, 
distancing us from our true objective: obtaining the state which gives us the maximum HTC with as few explored states as possible. 
This leads to the conclusion that, although we can miss the opportunity of learning a decent estimation of relevant state-action 
values, it is worth gathering states more slowly during the training as, just employing them, we can discover the optimal states 
too. A similar behavior will be also seen with policy-based algorithms.

Finally, Fig.~\ref{fig-double DQN trained}(c) shows that the highest HTC values are discovered early during the training. Again, 
this means that there is no need to reach a good policy in order to find the good states of our application, which suggests to 
train the $Q$-network with fewer states during more training time.

\subsection{Policy-based algorithms: REINFORCE} \label{sec-REINFORCE}

Now we focus on the analysis of the results obtained with REINFORCE \cite{Williams1992}, which is the most widely 
used policy-based RL algorithm. In this type of algorithms, the agent learns a policy function $\pi$, which in turn is used to produce 
actions and generate trajectories $\tau$ that maximize the objective $J(\tau)$. REINFORCE needs three components: 
(i) a parametrized policy, (ii) an objective to be maximized, like any other RL algorithm, and (iii) a method for updating 
the policy parameters. Concerning the parametrized policy, this is obtained with the help of deep neural networks that learn 
a set of parameters $\theta$. We denote the policy network as $\pi_\theta$ to emphasize that is parametrized by $\theta$. 
Framed in this way, the process of learning a good policy is equivalent to searching for a good set of values for $\theta$.

The objective that is maximized by an agent in REINFORCE is the expected return over all complete trajectories 
generated by an agent:
\begin{equation} \label{eq-J-reinforce}
	J(\pi_\theta) = \mathbb{E}_{\tau \sim \pi_\theta} \left[G(\tau) \right] = \mathbb{E}_{\tau \sim \pi_\theta} 
	\left[ \sum^{T}_{t=0} \gamma^t r_t \right] .
\end{equation}
Notice that the expectation is calculated over many trajectories sampled from a policy, that is, $\tau \sim \pi_\theta$. 
This expectation approaches the true value as more samples are collected, and it is specific of the policy $\pi_\theta$ used.

The final component of the algorithm is the policy gradient, which formally solves the following problem:
\begin{equation}
	\underset{\theta}{\rm max} \, J(\pi_\theta) = \mathbb{E}_{\tau \sim \pi_\theta} \left[G(\tau) \right] .
\end{equation}
To maximize the objective, we perform gradient ascent on the policy parameters $\theta$:
\begin{equation}
	\theta \leftarrow \theta + \alpha \nabla_{\theta} J(\pi_\theta) ,
\end{equation}
where $\alpha$ is the \emph{learning rate}, which controls the size of the parameter update. 
The term $\nabla_{\theta} J(\pi_\theta)$ is known as the \emph{policy gradient} and thanks to the policy gradient
theorem \cite{DeepRLbook,SuttonBarto} can be expressed as
\begin{equation} \label{eq-reinforce-nabla}
	\nabla_{\theta} J(\pi_\theta) =  \mathbb{E}_{\tau \sim \pi_\theta} \left[ \sum^{T}_{t=0} G_t(\tau) 
	\nabla_\theta \log \pi_\theta(a_t | s_t) \right] .
\end{equation}
Here, the term $\pi_\theta(a_t | s_t)$ is the probability of the action taken by the agent at time step $t$.
The action is sampled from the policy, $a_t \sim \pi_\theta(s_t)$. 

In practice, the REINFORCE algorithm numerically estimates the policy gradient using Monte Carlo sampling.  
Instead of sampling many trajectories per policy, one samples just one:
\begin{equation}
	\nabla_{\theta} J(\pi_\theta) \approx \sum^{T}_{t=0} G_t(\tau) \nabla_\theta \log \pi_\theta(a_t | s_t) .
\end{equation}
\begin{algorithm}[t] \label{alg-REINFORCE} 
    \caption{REINFORCE pseudocode \cite{Williams1992,DeepRLbook}.}
    \vspace{2mm}
    \KwIn{a differentiable policy parametrization $\pi_{\theta} : S \times \mathbb{R}^d \rightarrow A $}
    \KwIn{algorithm parameters. Initialize learning rate $\alpha > 0$ and discount rate $1 \geq \gamma \geq 0$}
    \KwOut{optimized policy $\pi_{\theta}$}
    
    Initialize arbitrarily $\pi_{\theta}$ weights, $\vec{\theta} \in \mathbb{R}^d$ 
    
    \For {episode}{
    
        Initialize $s_0 \neq terminal$

        Initialize $\nabla_{\theta}J(\pi_{\theta}) \leftarrow 0$

        \For {episode step}{
        
            Take $a_t$ following $\pi_{\theta}(a_t\vert s_t)$, observe $r_t$, $s_{t+1}$

            $s_t \leftarrow s_{t+1}$
            
        }

        \For {t = 0...T}{

            $G_t(\tau) \leftarrow \sum_{t'=t}^T\gamma^{t'-t}r_{t'}$
            
            $\nabla_{\theta}J(\pi_{\theta}) \leftarrow \nabla_{\theta}J(\pi_{\theta}) + 
            G_t(\tau)\nabla_{\theta}\log\pi_{\theta}(a_t\vert s_t) $
        
        }

        $\vec{\theta} \leftarrow \vec{\theta} + \alpha \nabla_{\theta}J(\pi_{\theta})$
    }
\end{algorithm}
\begin{figure*}[t]
\includegraphics[width=0.9\textwidth,clip]{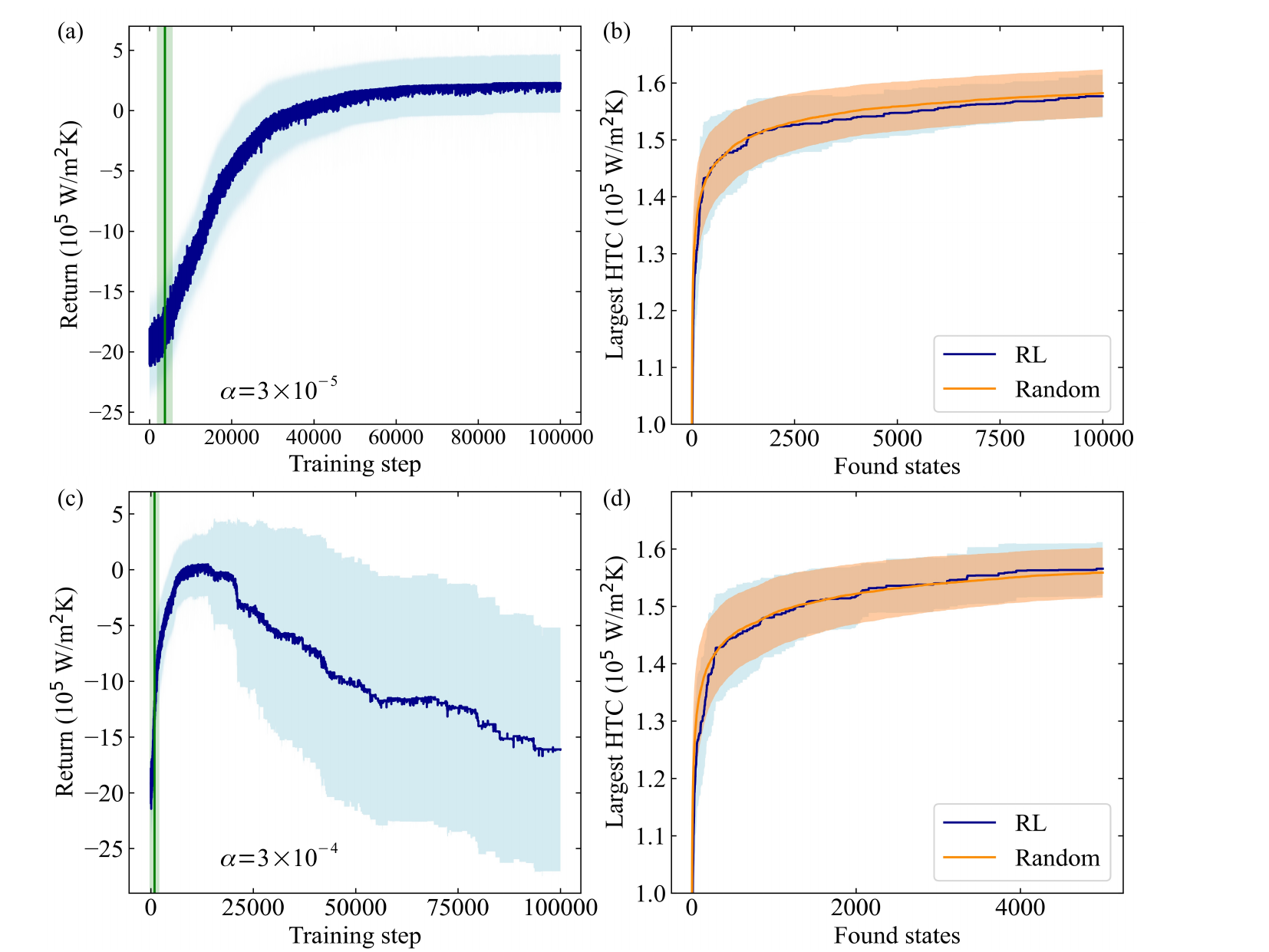}
\caption{Training of a REINFORCE algorithm. (a) Evolution of the return for the trajectories as training progresses. The 
vertical line and the green shaded area correspond to % the mean value and standard deviation of 
the training steps at which 
the highest HTC of the runs are found. Learning rate chosen as $\alpha = 3 \times 10^{-5}$. (b) Largest HTC discovered as 
a function of the number of found states for the evolution in (a). REINFORCE in blue versus a random search in orange. 
(c,d) Same as in (a,b), but with a learning rate $\alpha = 3 \times 10^{-4}$. In all panels the solid lines correspond to 
the mean value and the shaded areas to the standard deviations, as obtained in 40 independent runs for REINFORCE algorithm.}
\label{fig-reinforce}
\end{figure*}

The REINFORCE algorithm is summarized in pseudocode~\ref{alg-REINFORCE}. Let us emphasize how this is an episodic, 
\emph{on-policy} algorithm: every episode we start from the starting state $s_0$ in line 3 and collect a full trajectory 
$\tau = (s_0, a_0, r_0), \dots , (s_T , a_T , r_T)$ for an episode in lines 6 and 7. Then, in line 9 the return $G_t(\tau)$ 
is computed for each time step $t$ in the current trajectory, which is later used in line 10 to estimate the policy gradient 
along with this policy's action probabilities $\pi_\theta(a_t | s_t)$. In line 11, the sum of the policy gradients for all 
time steps calculated in line 10 is used to update the policy network parameters $\theta$, and the experiences are discarded.

It is known that the fact that the policy gradient is estimated sampling with a single trajectory typically leads to a high 
variance. One way to reduce this variance is to modify the returns by subtracting a suitable action-independent baseline as follows
\begin{equation}
	\nabla_{\theta} J(\pi_\theta) \approx \sum^{T}_{t=0} \left[ G_t(\tau) - b(s_t) \right] 
	\nabla_\theta \log \pi_\theta(a_t | s_t) .
\end{equation}
In our case we chose as baseline the mean return over the trajectory, namely, $b = \frac{1}{T}\sum^{T}_{t=0} G_t(\tau)$. 
Our baseline is state-independent, being a constant value for each trajectory, and centers the returns around 0. This 
enables faster learning, correcting any imbalance between positive and negative returns. Note that this REINFORCE-specific
baseline is independent of the baseline considered for the HTC values in the rewards.

Let us now describe our application of REINFORCE to our optimization problem. We first consider hyperparameters given in 
Table~\ref{tab:REINFORCE}. The corresponding evolution of the training for $10^{5}$ full episodes is shown in 
Fig.~\ref{fig-reinforce}(a), where we present the return of each episode as training progresses. These results corresponds 
to 40 independent runs. Notice that the return increases until it reaches a plateau, when the policy algorithm converges. 
The vertical line and the green shaded area correspond to the mean value and standard deviation of the training step
at which the highest HTC of the run is found. We can see that it was found in a much smaller number of steps compared 
to the total number of training steps needed for the convergence of the policy. This is a very similar result to what we 
found using DQN. In Fig.~\ref{fig-reinforce}(b) we show the comparison between the best HTC found by REINFORCE and by 
a random search, also for 40 different runs for REINFORCE algorithm and 1000 runs for the random algorithm. The random 
search algorithm is the same as in previous algorithms, so as to have a fair comparison. We can see how REINFORCE seems 
to behave in a very similar fashion as the random search, being unable to significantly outperform it.

To improve these results, we explored the same idea as in Double DQN, namely to increase the learning rate so it can learn 
with a reduced number of explored states. In Fig.~\ref{fig-reinforce}(c) we show the same as in Fig.~\ref{fig-reinforce}(a), 
but with a value of $\alpha = 3 \times 10^{-4}$, 10 times larger than before. As one can see, the training of the policy fails, 
with the return initially increasing but eventually decreasing abruptly, with a much larger standard deviation than before. 
In a regular RL problem, we would say that training has failed, and the learning rate is too high. However, if we look 
at Fig.~\ref{fig-reinforce}(d), same as in Fig.~\ref{fig-reinforce}(b) but with the new value for the learning rate, 
REINFORCE shows the same effect as in Double DQN: it is able to slightly outperform the random search algorithm in both the mean and
the area covered by the standard deviation. Moreover, the green shaded area in Fig.~\ref{fig-reinforce}(c) clearly demonstrates 
how the best examples are found with a really small number of training steps, even lower than in Fig.~\ref{fig-reinforce}(a), 
and the decay of the return is inconsequential to our goal. This is also analogous to our results in Double DQN.

We can further identify this improvement through the statistics of the runs, both from REINFORCE and the random search 
algorithm. Considering the $10^4$ new examples found in Fig.~\ref{fig-reinforce}(b), the random search algorithm is able to find 
one of the 5 configurations with highest HTC 57.0\% of the runs, with 17.3\% finding the actual maximum. On the other hand, 
REINFORCE is able to find one of the 5 highest HTC configurations 60\% of the runs, which is only slightly better than the random 
search, with 12.5\% finding the actual maximum, less than the random search. With this, we would conclude that the algorithm is worse 
than random search. But using the improved version in Fig.~\ref{fig-reinforce}(d), with a new restriction of 5000 new examples, 
which is half the amount of new states compared to Fig \ref{fig-reinforce}(b), the random search can 
only find one of the 5 configurations with highest HTC 34.9\% of the runs, with 9.3\% finding the actual maximum, and REINFORCE can 
find one of the 5 configurations with highest HTC 42.5\% of the runs, with 15\% finding the actual maximum, being able to surpass 
the random search results. However, the difference is not significant enough, and as such, REINFORCE is not the best option 
for solving our RL problem. The Monte Carlo sampling and the sample inefficiency of the algorithm prevent it from obtaining 
comparable or even better results than other algorithms. Still, its convergence is more easily guaranteed, albeit noisy, and its
implementation is more straightforward than value-based algorithms, so it can still be a candidate for other possible applications.

\subsection{Combined methods: Actor-Critic and PPO}

So far, we have discussed both policy-based and value-based algorithms. Now, we shall consider \emph{combined methods} which
learn two or more of the primary RL functions. To be precise, we shall discuss both \emph{Advantage Actor-Critic} (A2C) 
\cite{Konda1999} and \emph{Proximal Policy Optimization} (PPO) \cite{schulman2017}, which are among the most widely-used RL 
algorithms for a wide variety of applications. Actor-Critic algorithms receive their name from the two elements that compose 
them: an actor, which learns a parameterized policy like in REINFORCE; and a critic, which learns a value function to evaluate 
state-action pairs, becoming a learned reinforcement signal. In simple words, the foundations of actor-critic algorithms involve
trying to learn a value function to give the policy a more informative metric than just the rewards. When the learned 
reinforcement signal is the advantage function, the algorithm is called Advantage Actor-Critic. The advantage function is 
defined as follows:
\begin{equation}
	A^{\pi}(s_t,a_t) = Q^{\pi}(s_t,a_t) - V^{\pi}(s_t) ,
\end{equation}
thus describing how preferable an action would be compared to the average weighted by the policy in a particular state. This 
allows to rescale their values for all states and actions, similarly to a state-dependent baseline, and presents other useful
properties such as $\mathbb{E}_{a\in A}[A^{\pi}(s_t,a)] = 0$. The actor uses this signal in place of the estimate of the return 
from the REINFORCE algorithm, performing the same gradient ascent technique. That is, the actor performs the following policy 
optimization (setting $\pi_\theta \to \pi$):
\begin{equation}
	\nabla_{\theta}J(\pi) = \mathbb{E}_t [A_t^{\pi}\nabla_{\theta}\log \pi (a_t \vert s_t)].
\end{equation}
The critic is tasked with estimating this advantage function for all states and actions. In principle, this would imply being 
able to estimate both $Q^{\pi}(s,a)$ and $V^{\pi}(s)$, but there are methods that allow the estimation of $Q^{\pi}(s,a)$ through 
$V^{\pi}(s)$ over the trajectory, allowing us to only need to learn the latter. There are various ways to estimate the advantage 
function using these value functions. The first one is called $n$-step returns, in which we expand the definition of $Q^{\pi}(s,a)$ 
using the rewards obtained in the current trajectory for $n$ steps, and then take the $V$-value of the following one. That is, we 
expand:
\begin{eqnarray}
	Q^{\pi}(s_t,a_t) & = & \mathbb{E}_{\tau\sim\pi} \left[ r_t + \gamma r_{t+1} + \gamma^2 r_{t+2} + \cdots + \gamma^n r_{t+n}
     \right] 
    \nonumber \\ & & + \gamma^{n+1} V^{\pi} (s_{t+n+1}) \nonumber \\
    & \approx & r_t + \gamma r_{t+1} +  \gamma^2 r_{t+2} + \cdots + \gamma^n r_{t+n} \nonumber \\ 
    & & + \gamma^{n+1} V^{\pi} (s_{t+n+1}),
\end{eqnarray}
which assumes accurate $V^{\pi}(s)$ estimations. This leaves our bias-variance trade-off explicit: the rewards from the trajectory 
have high variance, while $V^{\pi}(s)$ is a biased estimation. Higher values of n present higher variance, so n should be chosen 
to balance these two effects. Another way to estimate the advantage is called \emph{Generalized Advantage Estimation} (GAE) 
\cite{schulman2018}. GAE calculates an exponentially-weighted average of all n-step advantages, intending to reduce the variance of 
the estimation while keeping the bias low. The expression is as follows
\begin{equation}
\begin{split}
	&A^{\pi}_{\rm GAE}(s_t,a_t) = \sum _{k=0}^{\infty}(\gamma\lambda)^k\delta_{t+k} \\
 &\text{where } \delta_t = r_t + \gamma V^{\pi}(s_{t+1}) - V^{\pi}(s_t),
\end{split}
\end{equation}
with $\lambda \in [0,1]$ controls the decay rate. A higher value introduces higher variance, up to a value of 1 which represents 
the Monte-Carlo estimate. A value of 0, on the other hand, computes the TD estimate of the returns. GAE is our estimation of choice 
for the advantage function, which we use for our physical problem. Finally, we need to obtain a way to estimate the $V^{\pi}(s)$ 
values for each state. Following the structure of value-based algorithms and the definition of advantage, we set the target 
of the critic network as
\begin{equation}
    V^{\pi}_{\text{tar}}(s_t) = A^{\pi}_{GAE}(s_t,a_t) + V^{\pi}(s_t),
\end{equation}
which we obtain in a similar manner to SARSA and Double DQN \cite{DeepRLbook}. Thus, the full A2C algorithm can be summarized in the
pseudocode~\ref{alg-A2C}.

\begin{algorithm}[t] \label{alg-A2C} 
    \caption{A2C pseudocode with GAE and MSE \cite{DeepRLbook}.}
    \vspace{2mm}
    \KwIn{a differentiable policy parametrization $\pi : S \times \mathbb{R}^{d_1} \rightarrow A $}
    \KwIn{a differentiable state-value function parametrization $\widehat{V} : S \times \mathbb{R}^{d_2} \rightarrow \mathbb{R}$}
    \KwIn{algorithm parameters. Initialize learning rate $\alpha_1 > 0$ and $\alpha_2 > 0$, GAE exponential weight 
    $1 \geq \lambda \geq 0$ and discount rate $1 \geq \gamma \geq 0$}
    \KwOut{optimized policy $\pi$ and improved $\widehat{V}$}
    
    Initialize arbitrarily $\pi$ weights, $\vec{\theta} \in \mathbb{R}^{d_1}$
    
    Initialize arbitrarily $\widehat{V}$ weights, $\vec{\omega} \in \mathbb{R}^{d_2}$
    
    \For {episode}{
    
        Initialize $s_0 \neq terminal$

        \For {episode step}{
        
            Take $a_t$ following $\pi(a_t\vert s_t)$, observe $r_t$, $s_{t+1}$

            $s_t \leftarrow s_{t+1}$
        }

        \For {t = 0...T}{

            Obtain $\widehat{V}(s_t)$

            Calculate $\widehat{A}^{\pi}_{\rm GAE}(s_t,a_t) = \sum _{k=0}^{T-t}(\gamma\lambda)^k\delta_{t+k}$

            Calculate $\widehat{V}^{\pi}_{\text{tar}}(s_t) = \widehat{A}^{\pi}_{\rm GAE}(s_t,a_t) + \widehat{V}^{\pi}(s_t)$
            
        }

        Obtain actor loss: $L_{\rm actor} = -\frac{1}{T}\sum_{t=0}^T\widehat{A}^{\pi}_{\rm GAE}(s_t,a_t)\log\pi(a_t\vert s_t)$
        
        Obtain critic MSE: $L_{\rm critic} = \frac{1}{T}\sum_{t=0}^T(\widehat{V}^{\pi}_{\text{tar}}(s_t)-\widehat{V}^{\pi}(s_t))^2$

        Update actor parameters: $\vec{\theta} \gets \vec{\theta} + \alpha_1 \nabla_{\theta}L_{\rm actor}$

        Update critic parameters: $\vec{\omega} \gets \vec{\omega} + \alpha_2 \nabla_{\omega}L_{\rm critic}$
        
    }
\end{algorithm}

Actor-critic algorithms present a series of issues too, some of the most common including performance collapse and sample 
inefficiency from being on-policy algorithms. PPO is one of the most popular algorithms designed to solve them, using a 
surrogate objective that ensures monotonic improvements and allows to reuse off-policy data samples. This new PPO objective 
replaces the original A2C objective, and could also be applied to REINFORCE. 

To understand this algorithm, we first need to consider that we are performing the search of optimal policies in the parameters 
space of $\Theta$, while the policies are sampled from the policy space $\Pi$. Thus, regular steps in the 
parameter space do not translate to regular steps in the policy space, where the optimal step size may vary depending on the 
local geometry, and might result in too big or too small policy steps. This is what eventually causes performance collapse. 

To avoid this issue, we consider a constraint to the change in the policy space. We define the distance in the objective 
between policies as \cite{schulman2017}
\begin{equation}
    J(\pi')-J(\pi) = \mathbb{E}_{\tau\sim\pi'} \left[ \sum_{t=0}^T\gamma^t A^{\pi}(s_t,a_t) \right],
\end{equation}
where $\pi$ is the original policy, which we used to calculate $A^{\pi}$, and $\pi'$ is the policy that we would obtain after 
the parameter update. This is a measure of the performance of the new policy, and our goal would be to maximize it. This new 
maximization problem ensures that there is always a monotonic positive improvement, since the worst possible result would be 
to let $\pi'=\pi$, without any modification to the policy.

Still, we cannot properly use this function as an objective function, because the expectation is performed sampling from the 
new policy, but the new policy is only available after the update that would require said new policy. To solve this issue, we
perform the sampling using the old policy, but including \emph{importance sampling} terms, the ratio between new and old policies.
Thus, the surrogate objective (renamed as $J(\pi')$) becomes an expectation over the current policy $\pi$, as desired:
\begin{equation}
    J(\pi') = \mathbb{E}_{\tau\sim\pi} \left[ \sum_{t=0}^T\gamma^t A^{\pi}(s_t,a_t)\frac{\pi'(a_t\vert s_t)}{\pi(a_t\vert s_t)}
    \right].
\end{equation}
Optimization using this objective is still gradient ascent, so this can become the new objective for the policy gradient function. 
Lastly, we need to check that the error of the estimation given by the approximation with importance sampling is not big enough 
to no longer fulfill the condition of always having a positive distance in the objective between policies. We know that, for 
sufficiently close policies, we can bind their error by their KL divergence \cite{achiam2017}. We only need to ensure that the 
policy improvement is bigger than this limit to accept a change. The application of this constraint can be quite straightforward: 
we simply need to constrain this KL divergence to be smaller than a given value, $\delta$. Thus, the problem becomes
\begin{equation}
\begin{split}
	& \max_{\theta}\mathbb{E} \left[ \frac{\pi'(s_t,a_t)}{\pi(s_t,a_t)}A^{\pi}(s_t,a_t) \right] \\
 &\text{ensuring } \mathbb{E}_t \left[ KL(\pi'(s_t,a_t)\vert\vert\pi(s_t,a_t)) \right] \leq \delta.
\end{split}
\end{equation}
We will solve this problem using PPO with clipped surrogate objective, a much simpler implementation doing away with the 
need to compute the KL divergence. For this, we define a hyperparameter $\epsilon$ to constrain the importance sampling terms. 
Thus, we constrain the objective between $(1-\epsilon)A_t^\pi$ and $(1+\epsilon)A_t^\pi$. The implementation is fairly simple, 
only needing to change the objective function from A2C in line 12 of pseudocode~\ref{alg-A2C} by this new function:
\begin{equation}
\begin{split}
    &J^{\rm CLIP}(\theta) = \mathbb{E}_t \left[ \min \left( \frac{\pi'(s_t,a_t)}{\pi(s_t,a_t)}
    A^{\pi}(s_t,a_t), \right. \right. \\ 
    & \left. \left. \text{clip} \left( \frac{\pi'(s_t,a_t)}{\pi(s_t,a_t)},1-\epsilon,1+\epsilon \right)
    A^{\pi}(s_t,a_t) \right) \right].
\end{split}
\end{equation}

Let us now describe our application of A2C and PPO to our optimization problem. Since these algorithms involve a larger
amount of hyperparameters, we made use of external packages to a more optimized implementation. The RL problems themselves were
implemented with the help of the library Stable-Baselines3 \cite{Raffin2021}, which features a set of reliable 
RL algorithms using PyTorch. For the optimization of hyperparameters, we used the library Optuna \cite{akiba2019}, an 
automatic hyperparameter optimization software framework. Still, fundamentally, there is no change in the algorithms themselves 
or their application. 

\begin{figure}[b]
\includegraphics[width=\columnwidth,clip]{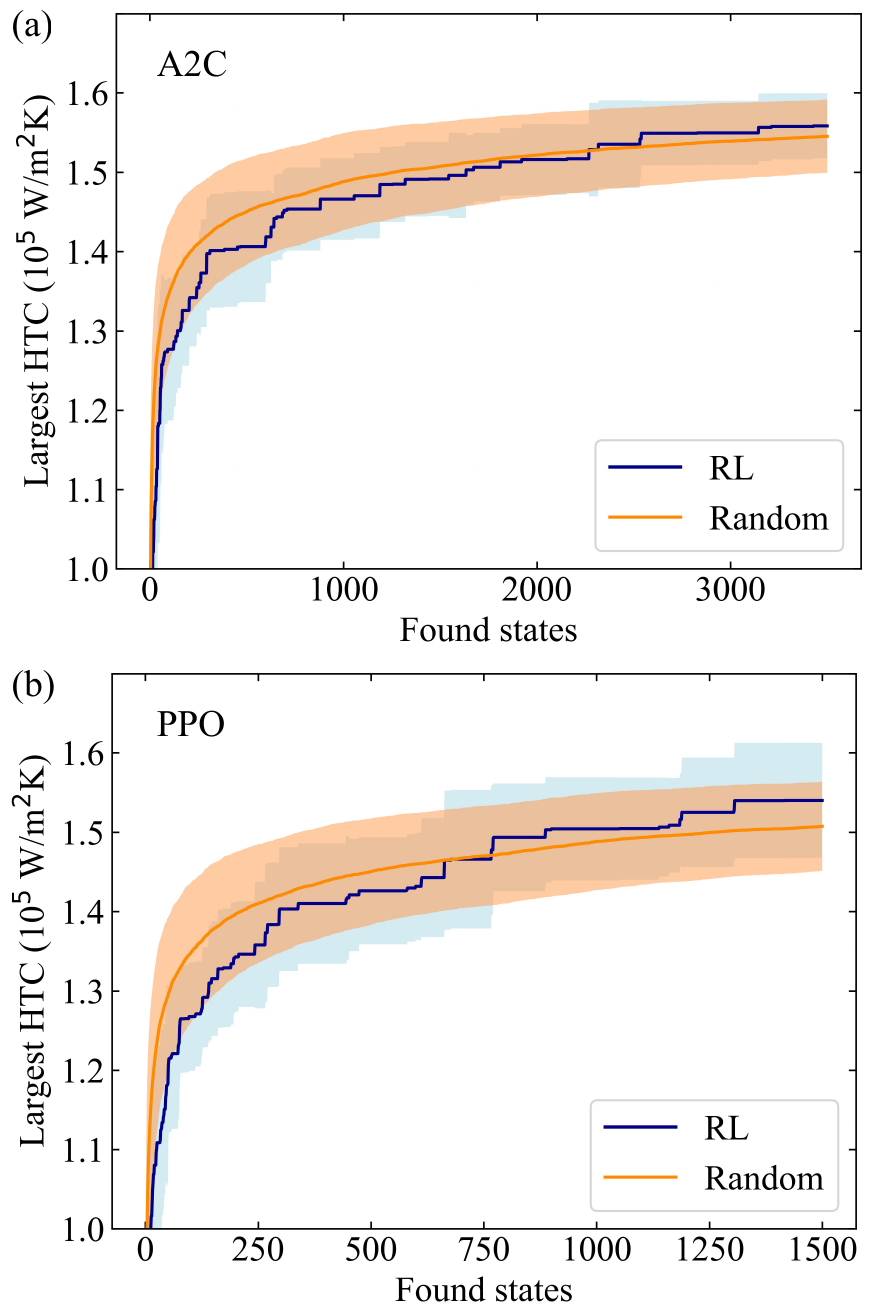}
\caption{Training of both A2C and PPO algorithms for our physical problem of interest. (a) Evolution of largest HTC found as new 
states are explored for the A2C algorithm. A2C in blue versus a random search in orange. (b) Same as in (a), but with 
PPO instead. In both panels the solid lines correspond to the mean value and the shaded areas to the standard deviations,
as obtained in 11 and 14 independent runs for A2C and PPO algorithms, respectively.}
\label{fig-A2C-PPO}
\end{figure}

The hyperparameters used in the implemented codes for A2C and PPO are detailed in Table~\ref{tab:A2C} and Table~\ref{tab:PPO}, 
respectively. Most of them were chosen and kept fixed from the beginning, with only the last 4 of their respective tables being
optimized by the Optuna hyperparameter search. The hyperparameters of said search are available in Table~\ref{tab:Optuna}.
The training results for both A2C and PPO are shown in Fig.~\ref{fig-A2C-PPO}, where we compare again the best 
state found by the algorithms and a random search, Fig.~\ref{fig-A2C-PPO}(a) for A2C and Fig.~\ref{fig-A2C-PPO}(b) for PPO. 
Both algorithms were trained for only $4000$ full episodes, since training with $10^5$ episodes like in REINFORCE yielded 
the exact same results. Following the format from the rest of the paper, the algorithm is presented in blue, in dark blue 
the average and in light blue the standard deviation, while the random search is presented in orange. The random search 
algorithm results are always obtained using 1000 runs, while for the RL algorithms, we only considered those that were able to 
perform a reasonable exploration of new found states, resulting in 11 for A2C and 14 in PPO. Note that this is similar to 
having considered a ``pruning" procedure over exploration, and does not consider how efficient those runs were in finding 
good states, only new ones.

Fig.~\ref{fig-A2C-PPO}(a) compares A2C and the random search for a total number of 3500 states found, much smaller than other 
algorithms. A2C is able to outperform the random algorithm slightly, in a similar fashion to REINFORCE, but it is not decidedly 
better. Fig.~\ref{fig-A2C-PPO}(b) compares PPO and the random search for a total number of 1500 states found, an even smaller 
number. PPO is able to outperform both the random search and the A2C algorithm with much smaller exploration. However, it also 
presents a wider standard deviation compared to A2C. 

We can more clearly describe these results by considering the statistics from the runs presented. For the 3500 found states 
in the study for A2C, the algorithm is able to find one of the 5 configurations with highest HTC 36\% of the runs, finding 
the best configuration 9.1\% of the runs. In comparison, the random search is only able to find one of the best 5 configurations 
in 26.2\% of the runs, and the best one in 6.8\% of them. For PPO, the results are even better than for A2C with a 
much lower number of states. PPO is able to find one of the best 5 configurations 29\% of the runs, with the best configuration 
21\% of the runs. Conversely, the random search can only find one of the top 5 configurations 12.3\% of the runs, and 3.2\%
the configuration with highest HTC. This makes PPO the best combined method for our purposes, and a strong candidate for a 
RL algorithm in these kind of problems.

\begin{figure}[t!]
\includegraphics[width=\columnwidth,clip]{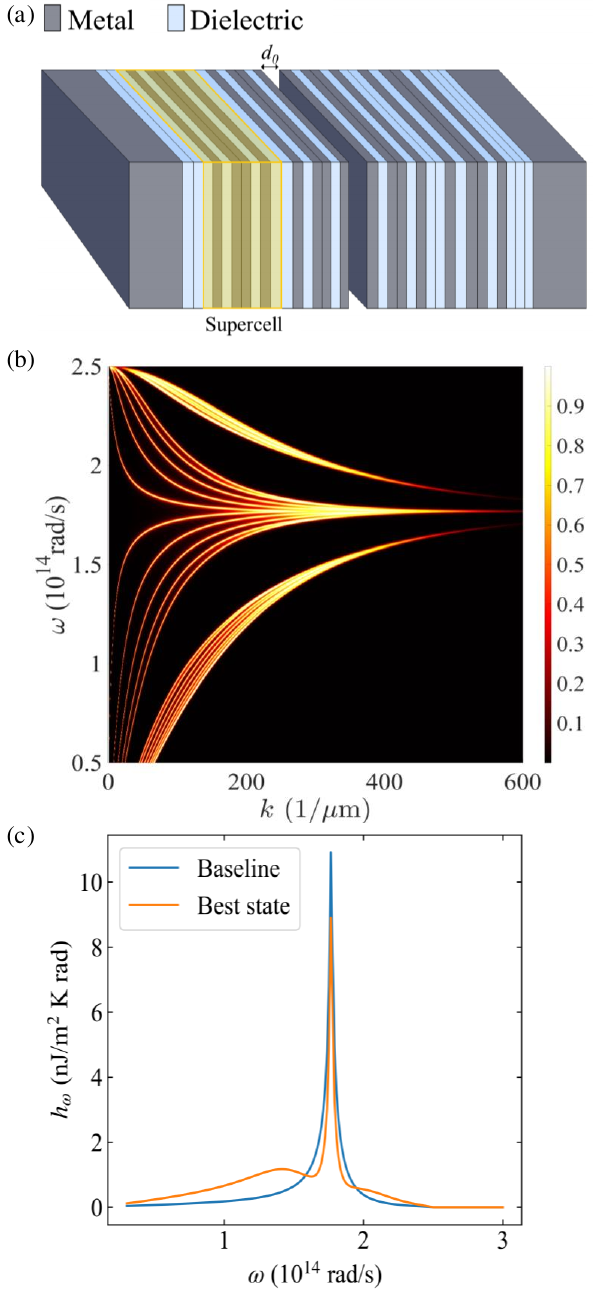}
\caption{Description of the best example found for the 16 layer problem. (a) Schematic structure of the best state found. 
A supercell structure is identified as a yellow shaded region on the left subsystem. 
(b) Transmission of evanescent waves as a function of the frequency ($\omega$) and the parallel wavevector ($k$) for the 
optimal structure shown in panel (a). (c) Comparison between the spectral HTC between the best state and the baseline of 
Fig.~\ref{fig-baseline}(b) at room temperature ($T=300$~K).}
\label{fig-best-state}
\end{figure}

\section{Discussion and conclusions} \label{sec-conclusions}

All algorithms employed in this work were able to identify, with differing success, the best configuration among the 
possibilities explored. This optimal 16 layer-configuration is shown in Fig.~\ref{fig-best-state}(a). Notice that it differs from 
the baseline configuration of Fig.~\ref{fig-baseline}(a), which corresponds to the periodic structure that one would propose 
based on physical intuition. The best state found by our algorithms does not seem to have any particular symmetry, although 
one can identify a periodic pattern forming a ``supercell", see Fig.~\ref{fig-best-state}(a). This pattern is repeated four 
times in the whole system when taking into account the gap -- which is two dielectric layers wide -- and both subsystems. 
The baseline configuration exhibits a HTC of $1.37 \times 10^5$ W/m$^2$ K, while the best state found has a HTC of $1.66 
\times 10^5$ W/m$^2$ K, a 21\% higher.

Figure~\ref{fig-best-state}(b) shows the transmission of evanescent waves as a function of the frequency $\omega$ and the 
parallel wavevector $k$ for the optimal configuration. The transmission pattern shows a series of narrow lines of values close 
to unity, which as explained in Sec.~\ref{sec-system} result from the hybridization of the SPPs that are formed in the
interfaces between the metallic and dielectric layers, similar to the baseline case in Fig.~\ref{fig-baseline}(b). We can 
notice some differences when comparing the two: while in the baseline case the lines all joined to form the same structure, 
in the optimal configuration we find a sizeable gap between some of the lines, likely stemming from the supercell structure mentioned 
above. This implies that their radiative heat transfer behavior is clearly different. In Fig.~\ref{fig-best-state}(c) we compare 
the spectral HTC $h_{\omega}$ of both the baseline (also shown in Fig.~\ref{fig-baseline}(c)) and the best state found. We can 
see that both present the main central resonant peak, but the one of the baseline is higher while the best state has a higher 
value in many other frequencies, with a notable secondary maximum at lower frequencies. The integration of these spectra over 
frequency yields the total HTC values mentioned at the end of the previous paragraph.

\begin{figure}[t]
\includegraphics[width=0.95\columnwidth]{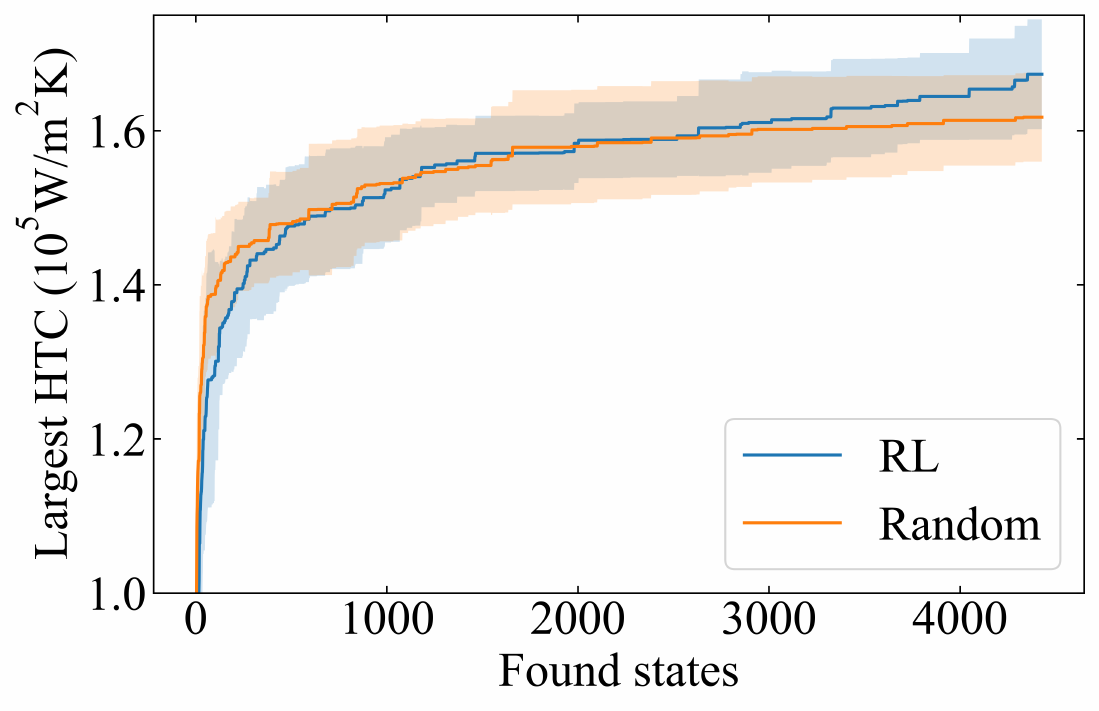}
\caption{Largest HTC discovered as a function of the number of found states in the problem with 24 active layers obtained 
with the Double DQN algorithm. We also present the results obtained with the random algorithm. The solid line corresponds 
to the mean value and the shaded area to the standard deviations, as obtained in 20 independent runs for both algorithms.}
\label{fig-24_layers}
\end{figure}

So far, we have focused our study on the case of 16 active layers. This case is complex enough for the illustration of the RL 
techniques while still being small enough that we only have $\sim$ 65 thousand possible configurations. Thus, we can directly 
find out which state has the highest HTC by simply analyzing all of them. To show that the RL methods reported here can also 
be applied to problems in which we do not know the solution beforehand, we consider now the case with 24 active layers, where 
the total number of states is $\sim 17$ million. We used the exact same techniques as before, but taking as baseline the perfectly 
periodic case with 24 layers instead of 16. We solved this problem using Double DQN, the value-based algorithm that showed the 
highest efficiency in the previous section, for 20 independent runs and 2500 training steps. The results for the largest HTC 
are shown in Fig.~\ref{fig-24_layers}. They were obtained using the optimal hyperparameters found for the 16-layer problem. 
Compared to the results discussed above, the RL algorithm surpasses the random algorithm after a few thousand explored states. 
In addition, notice that the random algorithm reaches a mean value of $1.617 \times 10^5$ W/m$^2$K, while the 
RL algorithm yields $1.673 \times 10^5$ W/m$^2$K, which in turn overcomes the value of the baseline state ($1.623 \times 10^5$ 
W/m$^2$K). This shows again that we can beat the physical intuition with the proper use of RL, reaching an state with a 
HTC value of $1.807 \times 10^5$ W/m$^2$K. Therefore, this analysis shows that the Double DQN algorithm can be successfully 
applied to a state space for which the computation of the HTC of all its states is not feasible in a reasonable amount of time.

So, in summary, we have shown in this work how RL can be used to tackle optimization problems in the context of radiative
heat transfer. As an illustration, we have addressed the maximization of the NFRHT between hyperbolic metamaterials made of 
a combination of metallic and dielectric layers. This problem is quite generic and contains the basic ingredients of most 
optimization problems in the field of thermal radiation. Our work demonstrates that these problems can be naturally 
formulated as a sequential decision-making problem and therefore, they are susceptible to be tackled with RL methods. It is
worth remarking that one could address inverse design problems in the same way by simply redefining the objective
function. 

In the physical problem studied in this work, we have shown that essentially all RL algorithms are able to find 
near optimal solutions, albeit with different efficiencies. In our case, we have found that Double DQN is the most efficient 
algorithm, with PPO also providing high quality results. While PPO finds the top 5 HTC values less frequently than Double 
DQN, it explores substantially fewer states and is still able to find the best ones much more reliably than the random 
algorithm. Therefore, both algorithms present their pros and cons in the application to our particular problem. In any case, 
we have provided a comprehensive guide on how to utilize in practice most of the key RL algorithms. Thus, we hope that our 
work will help other researchers to employ RL techniques as part of their toolkit for the investigation of optimization 
and inverse design problems in the context of radiative heat transfer and related topics.

The codes with the different RL algorithms used in this work are available from Ref.~\cite{DRL_RHT2024}. \\

\acknowledgments

E.O.M.\ thanks Komorebi AI Technologies team, with special mention to its founders, for their broad coding insights and 
discussions regarding some of the reported results. Furthermore, she acknowledges financial support by the Comunidad de 
Madrid and Komorebi AI Technologies under the Program ``Doctorados Industriales" (reference IND2022/IND-23536). 
J.J.G.E.\ was supported by the Spanish Ministry of Science and Innovation (Grant No.\ FPU19/05281). 
J.B.A.\ acknowledges funding from the Spanish Ministry of Science and Innovation (PID2022-139995NB-I00).
J.C.C.\ thanks the Spanish Ministry of Science and Innovation for financial support through Grant No.\
PID2020-114880GB-I00 and the ``Mar\'{\i}a de Maeztu" Programme for Units of Excellence in R\&D (CEX2023-001316-M). 

$^{\dagger}$ These authors contributed equally to this work.

\appendix

\section{Hyperparameters} \label{sec-hyperpara}

In Tables~\ref{tab-hyper-SARSA}-\ref{tab:PPO} we summarize the hyperparameters used in the different algorithms 
employed in this work. In particular, we describe its meaning and role.

\begin{table*}[h!]
    \centering
    \caption{Hyperparameters of SARSA algorithm.}
    \label{tab-hyper-SARSA}
    \begin{tabular}{|C{3cm}|C{1.5cm}|L{10cm}|}
        \hline 
        \textbf{Variable name} & \textbf{Value} & \textbf{Description} \\ 
        \hline
        hidden layers & 4 & Number of hidden layers of the neural network. \\
        hidden neurons & 64 & Number of neurons of each one of the hidden layers.\\
        \hline
        activation function & SELU & Activation function of the neurons. \\
        loss function & MAE & Loss function to evaluate the performance of the neural network.\\
        optimizer & Adam &  Optimizer during training.\\
        \hline
        $\gamma$ & $0.99$ & Discount rate parameter.\\
        episode length & 64 & Number of steps per episode. \\    
        \hline
        $\varepsilon$ & 1 & Initial value of epsilon parameter.\\
        $\varepsilon$ decay & $4 \times 10^{-5}$ & Value to reduce epsilon each train step.\\
        $\varepsilon$ minimum & $10^{-3}$ & Minimum reachable epsilon value. \\  
        $\alpha$ & $10^{-3}$ & Value of learning rate parameter.\\
        $\alpha$ decay & 0 & Value to reduce the learning rate each train step.\\
        \hline
        batch size & 32 & Number of experiences to process each train step. \\           
        \hline
    \end{tabular}
\end{table*}

\begin{table*}[h!]
    \centering
    \caption{Hyperparameters of Double DQN algorithm.}
    \label{tab-hyper-doubleDQN}
    \begin{tabular}{|C{3cm}|C{1.5cm}|L{10cm}|}
        \hline 
        \textbf{Variable name} & \textbf{Value} & \textbf{Description} \\ 
        \hline
        hidden layers & 4 & Number of hidden layers of the neural network. \\
        hidden neurons & 64 & Number of neurons of each one of the hidden layers.\\  
        \hline  
        activation function & SELU & Activation function of the neurons. \\
        loss function & MAE & Loss function to evaluate the performance of the neural network.\\
        optimizer & Adam & Optimizer during training.\\
        \hline
        $\gamma$ & $0.99$ & Discount rate parameter.\\
        episode length & $64$ & Number of steps per episode. \\     
        \hline        
        $\varepsilon$ & 1 & Initial value of epsilon parameter.\\
        $\varepsilon$ decay & $9 \times 10^{-5}$ & Epsilon decay each train step.\\
        $\varepsilon$ minimum & $10^{-3}$ & Minimum reachable epsilon value. \\      
        $\alpha$ & $10^{-4}$ & Value of learning rate parameter.\\
        $\alpha$ decay & 0 & Value to reduce the learning rate each train step.\\
        \hline    
        batch size & $32$ & Number of experiences to process each train step. \\    
        $B$ & $4$ & Number of batches processed each training step. \\     
        $K$ & $10^4$ & Memory size, maximum number of experiences stored in memory. \\ 
        $h$ & 4 & Number of new experiences added each train step. \\ 
        $F$ & $5 \times 10^3$ &  Number of train steps to update the target network.\\ 
        \hline
    \end{tabular}
\end{table*}

\begin{table*}[h!]
    \centering
    \caption{Hyperparameters of REINFORCE algorithm.}
    \label{tab:REINFORCE}
    \begin{tabular}{|C{3cm}|C{1.5cm}|L{8cm}|}
        \hline 
        \textbf{Variable name} & \textbf{Value} & \textbf{Description} \\ 
        \hline   
        hidden layers & $4$ & Number of hidden layers of the neural network.\\
        hidden neurons & $64$ & Number of neurons of each one of the hidden layers.\\
        \hline
        activation function & SELU & Activation function of the neurons. \\
        optimizer & Adam & Optimizer during training.\\
        $\alpha$ & $3 \times 10^{-5}$ & Value of learning rate parameter.\\
        \hline 
        $\gamma$ & $0.99$ & Discount rate parameter.\\
        episode length & $32$ & Number of experiences in an episode.\\
        n episodes & $10^5$ & Number of episodes in training.\\
        \hline  
    \end{tabular}
\end{table*}

\begin{table*}[h!]
    \centering
    \caption{Hyperparameters of A2C algorithm.}
    \label{tab:A2C}
    \begin{tabular}{|C{3cm}|C{1.8cm}|L{9cm}|}
        \hline 
        \textbf{Variable name} & \textbf{Value} & \textbf{Description} \\ 

        \hline
        
        hidden layers & $4$ & Number of hidden layers of the neural network, same for actor and critic networks.\\
        hidden neurons & $100$ & Number of neurons of each one of the hidden layers, same for actor and critic networks.\\
        
        \hline
        
        activation function & SELU & Activation function of the neurons. \\
        loss function & MAE & Loss function to evaluate the performance of the critic network.\\
        optimizer & Adam &  Optimizer during training.\\

        \hline
        
        episode length & $32$ & Number of experiences in an episode.\\
        n episodes & $10^5$ & Number of episodes in training.\\

        \hline

        gradient clip & True & Use of gradient clipping.\\
        vf coef & $1/2$ & Constant to weight the value function loss.\\
        use rms prop & False & Use RMSProp instead of Adam.\\
        
        \hline        
        
        $\gamma$ & $0.94645$ & Discount rate parameter.\\
        $\alpha$ & $3.53 \times 10^{-4}$ & Value of learning rate parameter.\\
        $\lambda$ & $0.997$ & Value of the GAE exponential factor.\\
        max grad & $1.285$ & Maximum gradient in the optimization step, gradient clipping.\\
        
        \hline
        
    \end{tabular}
\end{table*}

\begin{table*}[h!]
    \centering
    \caption{Hyperparameters of PPO algorithm.}
    \label{tab:PPO}
    \begin{tabular}{|C{3cm}|C{2cm}|L{9cm}|}
        \hline 
        \textbf{Variable name} & \textbf{Value} & \textbf{Description} \\ 

        \hline
        
        hidden layers & $4$ & Number of hidden layers of the neural network, same for actor and critic networks.\\
        hidden neurons & $100$ & Number of neurons of each one of the hidden layers, same for actor and critic networks.\\
        
        \hline
        
        activation function & SELU & Activation function of the neurons. \\
        loss function & MAE & Loss function to evaluate the performance of the critic network.\\
        optimizer & Adam &  Optimizer during training.\\

        \hline
        
        episode length & $32$ & Number of experiences in an episode.\\
        n episodes & $10^5$ & Number of episodes in training.\\
        $\epsilon$ & $0.2$ & Clipping range for the importance sampling terms.\\
        batch size & $32$ & Batch size for the PPO algorithm training.\\

        \hline

        gradient clip & True & Use of gradient clipping.\\
        vf coef & $1/2$ & Constant to weight the value function loss.\\
        
        \hline
        
        $\gamma$ & $0.99988$ & Discount rate parameter.\\
        $\alpha$ & $6.24 \times 10^{-5}$ & Value of learning rate parameter.\\
        $\lambda$ & $0.871$ & Value of the GAE exponential factor.\\
        max grad & $4.042$ & Maximum gradient in the optimization step, gradient clipping.\\
        
        \hline
        
    \end{tabular}
\end{table*}

\begin{table*}[h!]
    \centering
    \caption{Hyperparameters of Optuna search algorithm for A2C \& PPO.}
    \label{tab:Optuna}
    \begin{tabular}{|C{3cm}|C{2.2cm}|L{8.5cm}|}
        \hline 
        \textbf{Variable name} & \textbf{Value} & \textbf{Description} \\ 
        \hline
        n configs & $100$ & Number of hyperparameter configurations studied. \\
        n steps & $10^5$ & Number of experiences explored each configuration.\\
        \hline
        sampler & TPESampler & Sampler chosen to choose the next configuration.\\
        pruner & Median & Method of pruning a configuration choice.\\
        n startup & $5$ & Number of configurations before starting the pruner.\\
        n warmup & $3.33 \times 10^4$ & Experiences before checking for pruning.\\
        \hline
        $\gamma$ range & $[0.9,0.9999]$ & Range of exploration values for the $\gamma$ parameter. \\
        $\alpha$ range & $[10^{-6}, 10^{-3}]$ & Range of exploration values for the $\alpha$ parameter.\\
        $\lambda$ range & $[0.0, 1.0]$ & Range of exploration values for the $\lambda$ parameter.\\
        max grad range & $[0.5, 5.0]$ & Range of exploration values for the max grad parameter.\\
        \hline
    \end{tabular}
\end{table*}

\bibliography{bibliography}

\end{document}